\documentclass[aps,pra,reprint,groupedaddress]{revtex4-1}
\usepackage{graphicx}
\usepackage{dcolumn}
\usepackage{bm}
\usepackage{color}
\usepackage{amsmath, amsthm, amssymb, amsfonts}
\usepackage{setspace}

\begin{document}

\title{Nonreciprocal microwave transmission based on Gebhard-Ruckenstein hopping}
\author{Shumpei Masuda$^1$}
\email{masulas@tmd.ac.jp}
\author{Shingo~Kono$^2$}
\author{Keishi Suzuki$^2$}
\author{Yuuki Tokunaga$^3$}
\author{Yasunobu Nakamura$^{2,4}$}
\author{Kazuki Koshino$^1$}
\affiliation{$^1$College of Liberal Arts and Sciences, Tokyo Medical and Dental University,
Ichikawa, Chiba 272-0827, Japan}
\affiliation{$^2$Research Center for Advanced Science and Technology, The University of Tokyo, Meguro-ku, Tokyo 153-8904, Japan}
\affiliation{$^3$NTT Secure Platform Laboratories, NTT Corporation, Musashino 180-8585, Japan}
\affiliation{$^4$RIKEN Center for Emergent Matter Science, Wako, Saitama 351-0198, Japan}

\date{\today}

\begin{abstract}
We study nonreciprocal microwave transmission based on the Gebhard-Ruckenstein hopping.
We consider a superconducting device that consists of microwave resonators and a coupler.
The Gebhard-Ruckenstein hopping between the resonators gives rise to a linear energy dispersion which manifests chiral propagation of microwaves in the device.
This device functions as an on-chip circulator with a wide bandwidth when transmission lines are attached.
\end{abstract}

\maketitle

\section{Introduction}
As erythrocytes transport oxygen from the lungs to the body tissues, microwaves can carry energy and information between electromagnetic components in superconducting circuits \cite{Blais2004,Wallra2004,Astafiev2007,Hofheinz2009,Inomata2016,Kuan2017}, which provide a promising platform for quantum information processing~\cite{Nakamura1999,Blais2004,Wallra2004,Astafiev2007,Majer2007,Sillanpaa2007,Wolf2009,Astafiev2010,Devoret2013,Kelly2015,Ofek2016,Koshino2017}.
Many of the quantum information processing schemes and the superconducting quantum optics experiments require routing of microwaves in a cryostat. 
Therefore, cryogenic circulators are indispensable tools, and the loss at each circulator is detrimental especially for quantum information processings. 
This motivates the growing body of experimental and theoretical works devoted to lossless on-chip microwave circulators, which possibly replace the commercial ferrite circulators.

Various principles for achieving on-chip microwave circulators, as well as their practical designs,  have been proposed  
~\cite{Koch2010,Kamal2011,Nunnenkamp2011,Hafezi2012,Kerckhoff2015,Metelmann2015,Ranzani2015,Walter2016}.
Several types of the devices with the nonreciprocal transmission of microwaves have been implemented such as, electrically driven nonreciprocity on a silicon chip~\cite{Lira2012}, a circulator based on a Josephson circuit~\cite{Silwa2015}, a fiber-integrated quantum optical circulator operated by a single atom~\cite{Scheucher2016}, chiral ground-state currents of interacting photons in three qubits 
based on a synthetic magnetic field~\cite{Roushan2017}, on-chip nonreciprocal current based on a combination of frequency conversion and delay~\cite{Chapman2017Nov} and the ones based on optomechanical circuits~\cite{Peterson2017July,Bernier2017Sep,Barzanjeh2017Oct}.
The nonreciprocal signal routing in photonic resonator lattice systems has been also studied \cite{Fang2012,Metelmann2017}.



In this paper, we investigate the microwave response of a system of the coupled resonators
with the Gebhard-Ruckenstein (GR) hopping~\cite{Gebhard1992,Kuramoto2009}.
The GR hopping gives rise to a linear energy dispersion, 
which manifests the chiral hopping of cavity photons. 
This system functions as a circulator when transmission lines are attached to some (three or more) of the resonators.
This paper is organized as follows.
In Sec.~\ref{Gebhard-Ruckenstein hopping} 
we  discuss the property of the system with GR hopping.
In Sec.~\ref{Model} we introduce our circulator based on the GR hopping.
In Sec.~\ref{Numerical results} 
we analyze the microwave response of the system and demonstrate the robustness of the routing efficiency of the circulator.
Section~\ref{conclusion and discussion} is devoted to conclusion.

\section{Gebhard-Ruckenstein hopping}
\label{Gebhard-Ruckenstein hopping}
\begin{figure}[h!]
\begin{center}
\includegraphics[width=2.7cm]{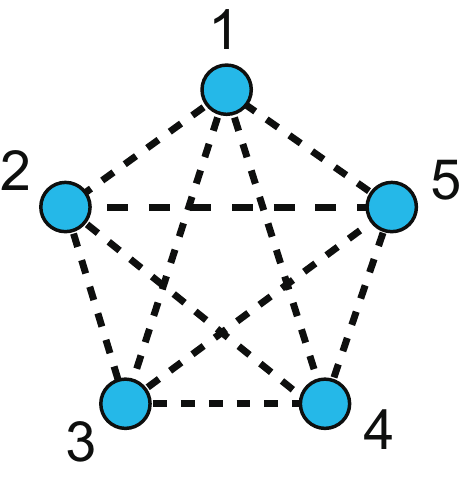}
\end{center}
\caption{
Schematic of a GR cluster for $N=5$.  The dashed lines represent the coupling between the sites.
}
\label{modelGR_11_17_17}
\end{figure}

In this paper we propose a method to route microwaves based on the GR hopping between the resonators.
To illustrate the property of GR hopping we introduce a system, which we call GR cluster, consisting of $N$ bosonic sites (resonators) with the GR hopping as depicted in Fig.~\ref{modelGR_11_17_17} .

It is known that the GR hopping gives rise to a linear energy dispersion, which manifests a chiral current in the system.
The GR hopping was also used to model the helical current on the edge of a two-dimensional topological insulator~\cite{masuda2012}.
The Hamiltonian of the GR cluster is represented as
\begin{eqnarray}
H_{\rm cluster}^{(\rm GR)}  = \sum_{m,n} \hbar \eta_{m,n}c_m^{\dagger}c_n,
\label{Hring}
\end{eqnarray}
with the bosonic annihilation operator $c_m$ for site~$m$ and the coupling constant between site~$m$ and site~$n$ given by
\begin{eqnarray}
\eta_{m,n} =  \eta_{n,m}^{\ast} = 
\begin{cases}
\frac{i\pi\eta_0(-1)^{n-m}}{N\sin\frac{\pi(n-m)}{N}}  & (n\ne m) \\
0 & (n=m) \\
\end{cases},
\label{tlm}
\end{eqnarray}
where $\eta_0$ is a real constant.
The single-particle eigenstates and their eigenenergies are represented as
\begin{eqnarray}
|\nu\rangle &=& \sum_{m=1}^N \phi_\nu(m) |m\rangle,\nonumber\\
E_\nu &=& \hbar \eta_0  k_\nu,
\end{eqnarray}
respectively, with the wave function $\phi_\nu(m) =\frac{1}{\sqrt{N}} \exp(i k_\nu m)$,
where $\nu=-(N-1)/2 ,\cdots, (N-1)/2$ for odd $N$, and $\nu=-N/2 ,\cdots, N/2-1$ for even $N$.
Here, $|m\rangle$ represents the state in which the particle (photon) is localized at site $m$.
$k_\nu$ is given by 
$k_\nu={2\pi \nu}/{N}$ for odd $N$, and
$k_\nu={2\pi (\nu+1/2)}/{N}$ for even $N$.

\begin{figure}[h!]
\begin{center}
\includegraphics[width=8cm]{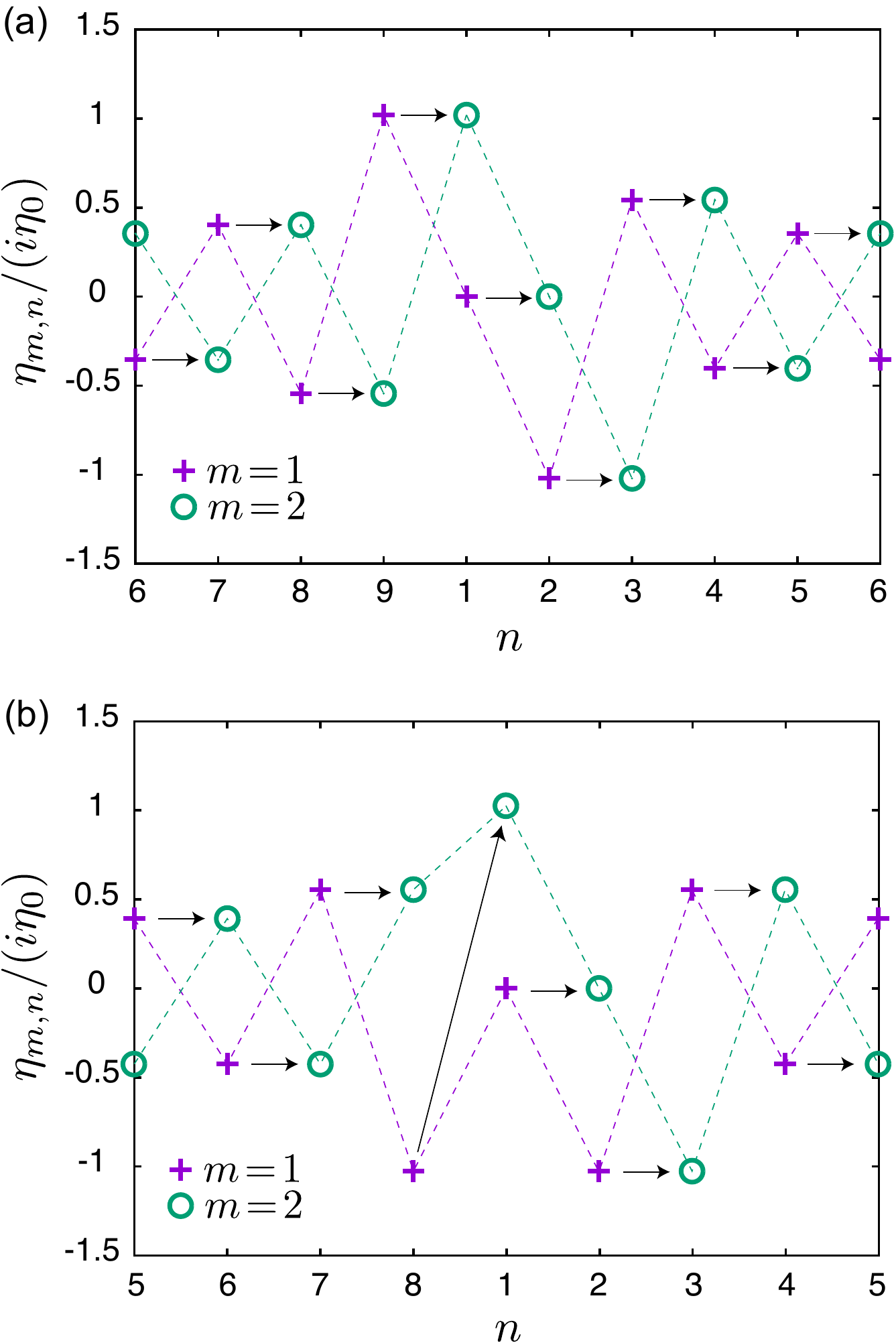}
\end{center}
\caption{
Coupling constant in the GR model.
$\eta_{m,n}$ for $m=1,2$ are shown for $N=9$ and $8$ in panels (a) and (b), respectively.
Note that the rightmost and leftmost values in the panels are identical.
The dashed lines are the guide to the eyes.
The arrows in panel (a) indicate the cyclicity of the Hamiltonian elements for $N=9$, and the arrows in panel~(b) indicate the lack of the cyclicity for $N=8$.
}
\label{hami_10_3_17}
\end{figure}

Figures~\ref{hami_10_3_17}(a) and \ref{hami_10_3_17}(b) show 
$\eta_{m,n}/(i\eta_0)$ for $N=9$ and $8$, respectively. 
The sign of $\eta_{m,n}$ changes alternately with respect to the site number~$n$ (except for $n=N$ for even $N$).
Note that the coupling constants are cyclic for odd $N$, that is, $\eta_{m,n}=\eta_{m+j,n+j}$ for integer $j$, where the indices are understood modulo $N$.
In contrast, this does not hold for even $N$: $\eta_{N,1}=-\eta_{1,2}$ (see Fig.~\ref{hami_10_3_17}(b)).

\begin{figure}[h!]
\begin{center}
\includegraphics[width=8.5cm]{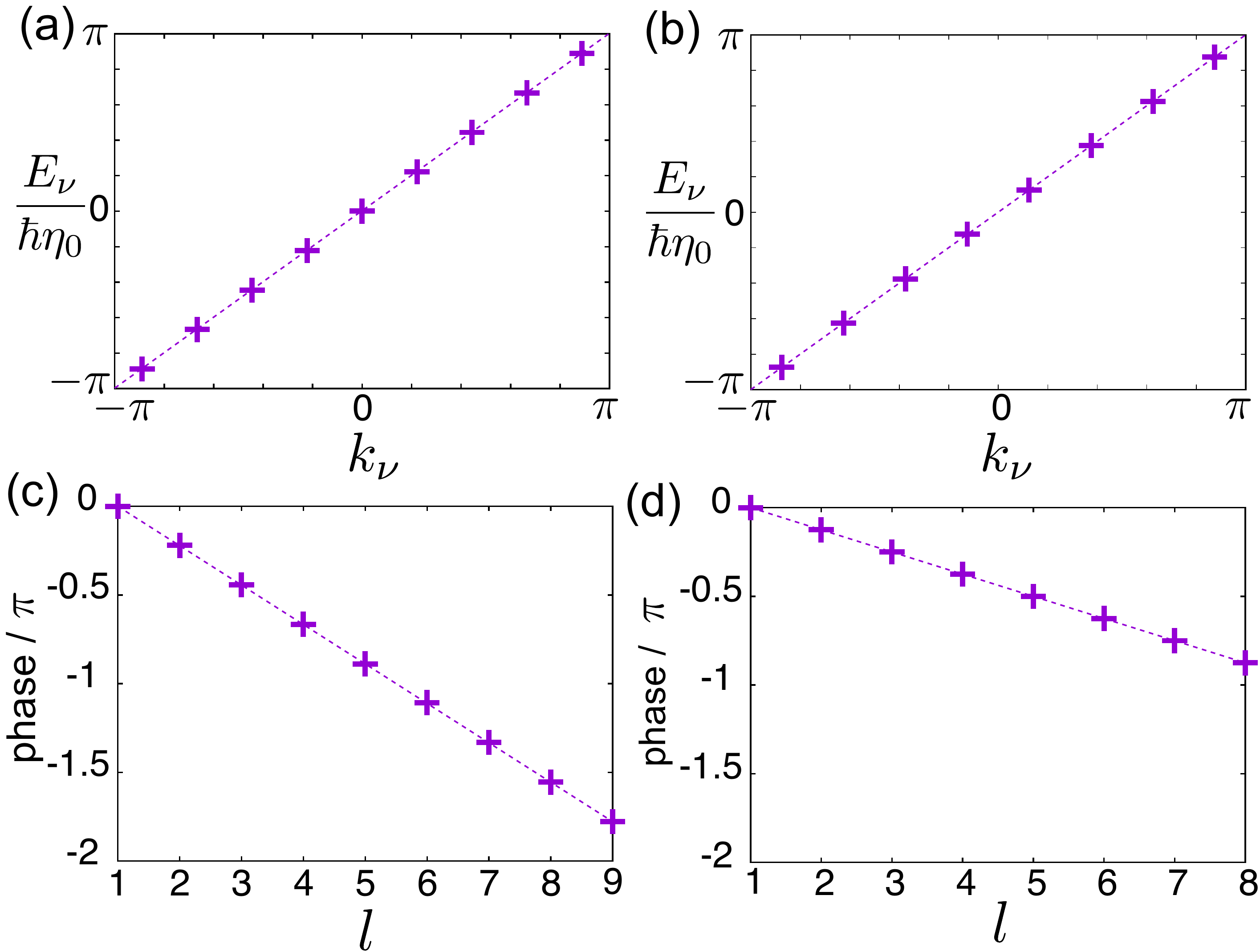}
\end{center}
\caption{
Eigenenergies of $H_{\rm cluster}^{\rm(GR)}$ as the functions of $k_\nu$ for (a) $N=9$ and (b) $N=8$.
Panels~(c) and (d) show the phase of wave function $\phi_{-1}(l)$ relative to that of $\phi_{-1}(1)$ for $N=9$ and 8, respectively.
}
\label{E_wf_10_3_17}
\end{figure}

Figures~\ref{E_wf_10_3_17}(a) and \ref{E_wf_10_3_17}(b) plot the eigenenergies as the functions of wave number.
Figures~\ref{E_wf_10_3_17}(c)  and \ref{E_wf_10_3_17}(d) show the phase of the wave function 
of the fourth lowest level relative to that of site~1.
The phase of the wave function of the eigenstates changes approximately $2\pi \nu$ from site~1 to site~$N$ for odd $N$, while it changes approximately $2\pi (\nu+1/2)$ for even $N$ as shown in Figs.~\ref{E_wf_10_3_17}(c) and \ref{E_wf_10_3_17}(d). 

\begin{figure}[h!]
\begin{center}
\includegraphics[width=7.5cm]{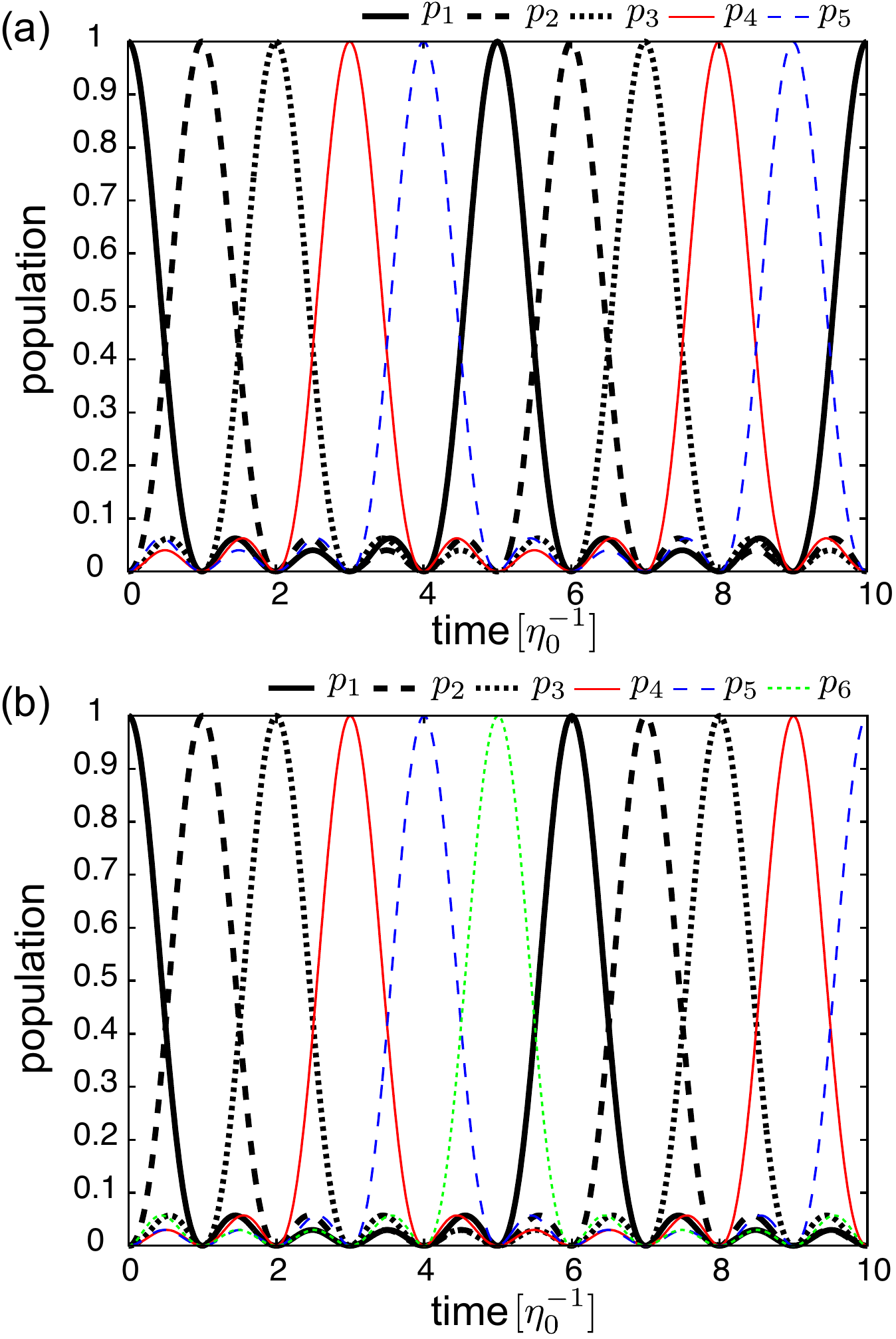}
\end{center}
\caption{
Time-evolution of the population at each site for the systems with (a)~$N=5$ and (b) $N=6$.
$p_i$ denotes the population of site~$i$.
}
\label{p_com_11_17_17}
\end{figure}

In Fig.~\ref{p_com_11_17_17}, we observe the dynamics of a particle (photon) in the GR clusters.
The initial state is set as $|\Psi\rangle=|1\rangle$.
Figures~\ref{p_com_11_17_17}(a) and~\ref{p_com_11_17_17}(b) show the time-evolution of the population at each site for the systems with $N=5$ and $6$, respectively.
The figures clearly show the chiral population transfer in the GR clusters.

\section{System}
\label{Model}
\subsection{Circulator based on GR hopping}
\label{Transmission via GR cluster}
\begin{figure}[h!]
\begin{center}
\includegraphics[width=4cm]{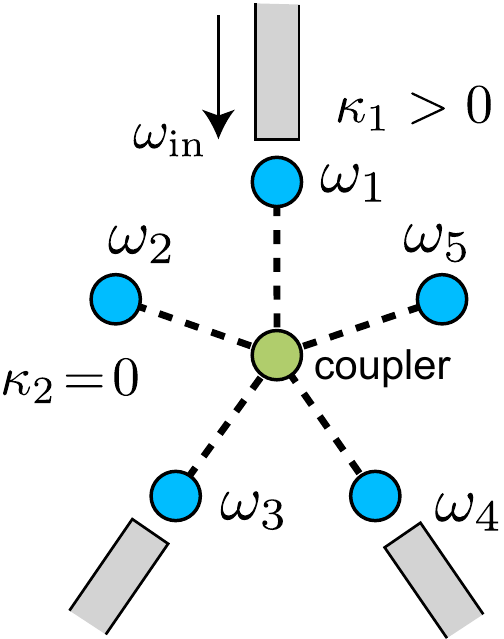}
\end{center}
\caption{
Schematic of a circulator with $N=5$.
The input field is applied through one of the transmission lines.
The blue circles, the green circle and gray lines represent the resonators, the coupler, and the transmission lines, respectively.
The dashed lines represent the coupling between the resonators.
Transmission lines are uncoupled to resonators~2 and~5 in this example.
}
\label{model_6_6_17}
\end{figure}

In this section, we propose a way of routing microwaves based on the GR hopping between resonators.
As depicted in Fig.~\ref{model_6_6_17}, our system consists of the $N$ resonators with different resonance frequencies, a coupler of the resonators and  transmission lines coupled to some of the resonators. 
We assume that each resonator is coupled to the other resonators with time-dependent coupling strength. (Physical realization is discussed in Appendix~\ref{Effective photon Hamiltonian}.)
The Hamiltonian of the system is given by 
\begin{eqnarray}
\mathcal{H} &=& \mathcal{H}_{\rm cluster} + \mathcal{H}_{\rm damp}, \nonumber \\
\mathcal{H}_{\rm cluster} &=& \sum_{m=1}^N \hbar \omega_m a_m^\dagger a_m
+ \sum_{m,n(\ne m)}^N \hbar g_{m,n}(t) a_m^{\dagger}a_n,\nonumber \\ 
\mathcal{H}_{\rm damp} &=& \hbar \sum_{m=1}^N \int dk \Big{[} v k b_{m,k}^\dagger b_{m,k} \nonumber \\ 
&& +
 \sqrt{\frac{v\kappa_m}{2\pi}}(a_m^\dagger b_{m,k} +  b_{m,k}^\dagger a_m) \Big{]}.
\label{H_6_6_17}
\end{eqnarray}
Here, $\mathcal{H}_{\rm cluster}$ describes the $N$ coupled resonators, and $\mathcal{H}_{\rm damp}$ describes the interactions between the transmission lines and the resonator modes.
$a_m$ and $b_{m,k}$ are the annihilation operators of the mode of resonator~$m$ and the mode of the transmission line~$m$ with wave number $k$, respectively.
We refer to the transmission line attached to resonator~$m$ as transmission line $m$.
$\omega_m$ is the resonance frequency of the mode of resonator~$m$,
$v$ is the microwave velocity in the transmission lines,
$\kappa_m$ is the decay rate of a photon from resonator~$m$ into transmission line $m$.
Note that transmission lines are not attached to some of the resonators. 
For example, in Fig.~\ref{model_6_6_17}, $\kappa_2=\kappa_5=0$.


The coupling between resonators~$n$ and~$m$ is modulated periodically in time as
\begin{eqnarray}
g_{m,n}(t) = 2\overline{g}_{m,n} \cos\Big{[}(\omega_m - \omega_n) t + \theta_{m,n} \Big{]},
\label{g_6_6_17}
\end{eqnarray}
where $\overline{g}_{m,n}$ and $\theta_{m,n}$ are time independent, and $g_{n,m}(t) = g_{m,n}^\ast(t)$.
The frequency of $g_{m,n}$ in Eq.~(\ref{g_6_6_17}) was set so that resonator~$m$ couples to resonator~$n$.
In the rotating frame the coupling strength between resonators~$m$ and~$n$
becomes independent of time and
is given by 
$\overline{g}_{m,n} e^{-i\theta_{m,n}}$ (see Appendix \ref{Smatrix} for details).
Thus, the phase of the coupling strength can be tuned {\it in-situ} via $\theta_{m,n}$.

Hereafter, we restrict ourselves to the case in which the coupling between the resonators are of the GR type and three of the resonators are coupled to the transmission lines (see Fig.~\ref{model_6_6_17}).
We set $\overline{g}_{m,n}$ and $\theta_{m,n}$ in Eq.~(\ref{g_6_6_17}) as
\begin{eqnarray}
\overline{g}_{m,n} &=& 
\frac{\pi \eta_0(-1)^{n-m}}{N\sin\frac{\pi (n-m)}{N}},\nonumber\\
\theta_{m,n} &=& - \frac{\pi}{2},
\end{eqnarray}
so that the coupling strengths between the resonators, $\overline{g}_{m,n} e^{-i\theta_{m,n}}$, become the GR type.
As shown in Sec.~\ref{Numerical results} this system has the non-reciprocal transmission property and functions as a circulator.

We comment here on the relation between our circulator and that studied previously.
The $N=3$ case in Eq.~(\ref{H_6_6_17}) was studied in Ref.~\citenum{Roushan2017} to show that the unit routing efficiency ($|S_{12}|=|S_{23}|=|S_{31}|=1$) is achieved when
\begin{eqnarray}
g_{m,n} &=& g, \nonumber \\
\kappa_m &=& 2g,  \nonumber \\
\theta_{1,2} &=& \theta_{2,3} = \theta_{3,1} =  \pi/2. 
\label{conditions_7_5_17}
\end{eqnarray}
Note that $\theta_{n,m}= - \theta_{m,n}$.
Our circulator based on the GR hopping for $N=3$ with $\eta_0 =3g\sin(\pi/3)/\pi$ and $\kappa_m = 2g$ 
satisfies Eq.~(\ref{conditions_7_5_17}) and reduces to the same system.

\section{Results}
\label{Numerical results}
Wide bandwidth is a desirable property of a circulator.
We examine the stability of the routing efficiency against detuning of the incident microwaves, assuming that the strength of the coupling between resonators is limited.
We consider the case in which every transmission line is coupled to a resonator with the same strength, $\kappa$, and the amplitude of the nearest neighbor hopping of the GR cluster, which is the largest, is fixed to be $g$ unless it is stated that we consider other cases.
Hereafter $g$ is used as the unit of detuning.
In the following, we first present the results for odd $N$, and then show the results for even $N$.

\subsection{$S$-matrix}
We consider the injection of continuous microwave from one of the transmission lines,
transmission line~$p$.
As shown in Appendix \ref{Smatrix}, the transmission and reflection coefficients of the microwave are given by the $S$-matrix elements represented as  
\begin{eqnarray}
S_{p,m} = \delta_{p,m}   - \sqrt{{\kappa_p\kappa_m}} [\mathcal{G}^{-1}]_{m,p},
\label{S_7_1_17}
\end{eqnarray}
where $[\mathcal{G}^{-1}]_{m,p}$ is the element of matrix $\mathcal{G}^{-1}$. 
The elements of matrix $\mathcal{G}$ are given by
\begin{eqnarray}
\mathcal{G}_{m,n}= \begin{cases}
\kappa_m/2 - i\Delta \omega & (n=m), \\
i \overline{g}_{m,n} e^{-i\theta_{m,n}} & (n\ne m),
\end{cases}
\label{G_9_5_17}
\end{eqnarray}
where $\Delta \omega (\equiv\omega_{\rm in}-\omega_p)$ is the detuning of the incident microwave, and 
$\omega_{\rm in}$ is the frequency of the incident microwave.

\subsection{three-resonator system}
\label{three-resonator system}
\begin{figure}[h!]
\begin{center}
\includegraphics[width=7.7cm]{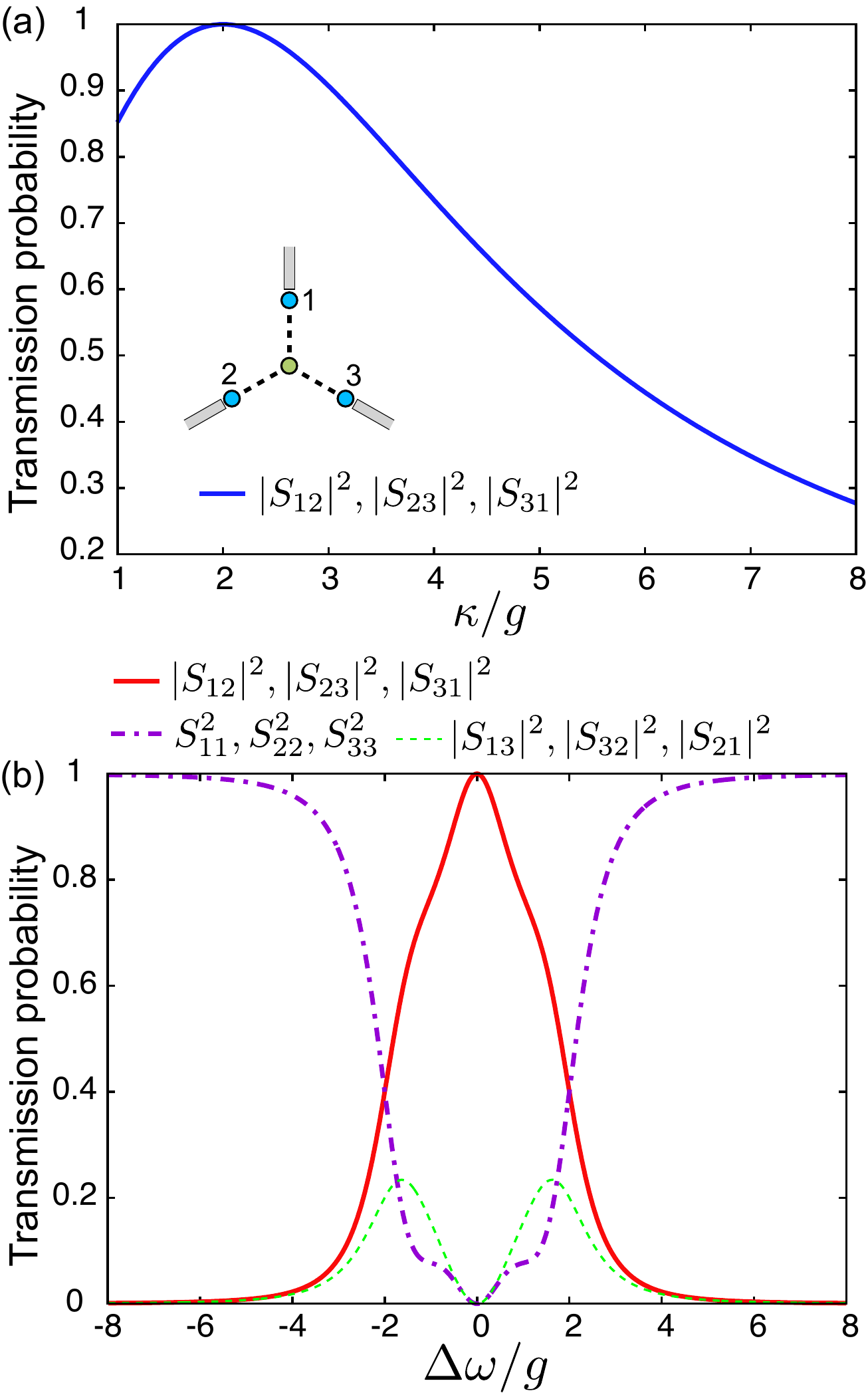}
\end{center}
\caption{
Circulation properties for the case of $N=3$ and $\kappa_{1,2,3}=\kappa$.
(a) Dependence of the forward transmission probabilities ($|S_{12}|^2,\ |S_{23}|^2$ and $|S_{31}|^2$) on $\kappa$ for $\Delta\omega=0$.
The inset shows the system configuration.
(b) Dependence of the forward and backward transmission and reflection probabilities on $\Delta \omega$ for $\kappa=2g$.
}
\label{S2_omega_kappa_3site_2_25_18}
\end{figure}

For the reference, we examine the routing efficiency of the three-resonator system.
Figure~\ref{S2_omega_kappa_3site_2_25_18}(a) shows the dependence of  the forward transmission probabilities (counter clockwise) on $\kappa$ for $N=3$ without detuning.
The forward transmission probabilities become unity at $\kappa=2g$.
Figure~\ref{S2_omega_kappa_3site_2_25_18}(b) shows the dependence of the forward and backward transmission and reflection probabilities on detuning $\Delta \omega$ with $\kappa=2g$. 

\subsection{five-resonator system}
\label{five-resonator system}
\begin{figure}[h!]
\begin{center}
\includegraphics[width=7.7cm]{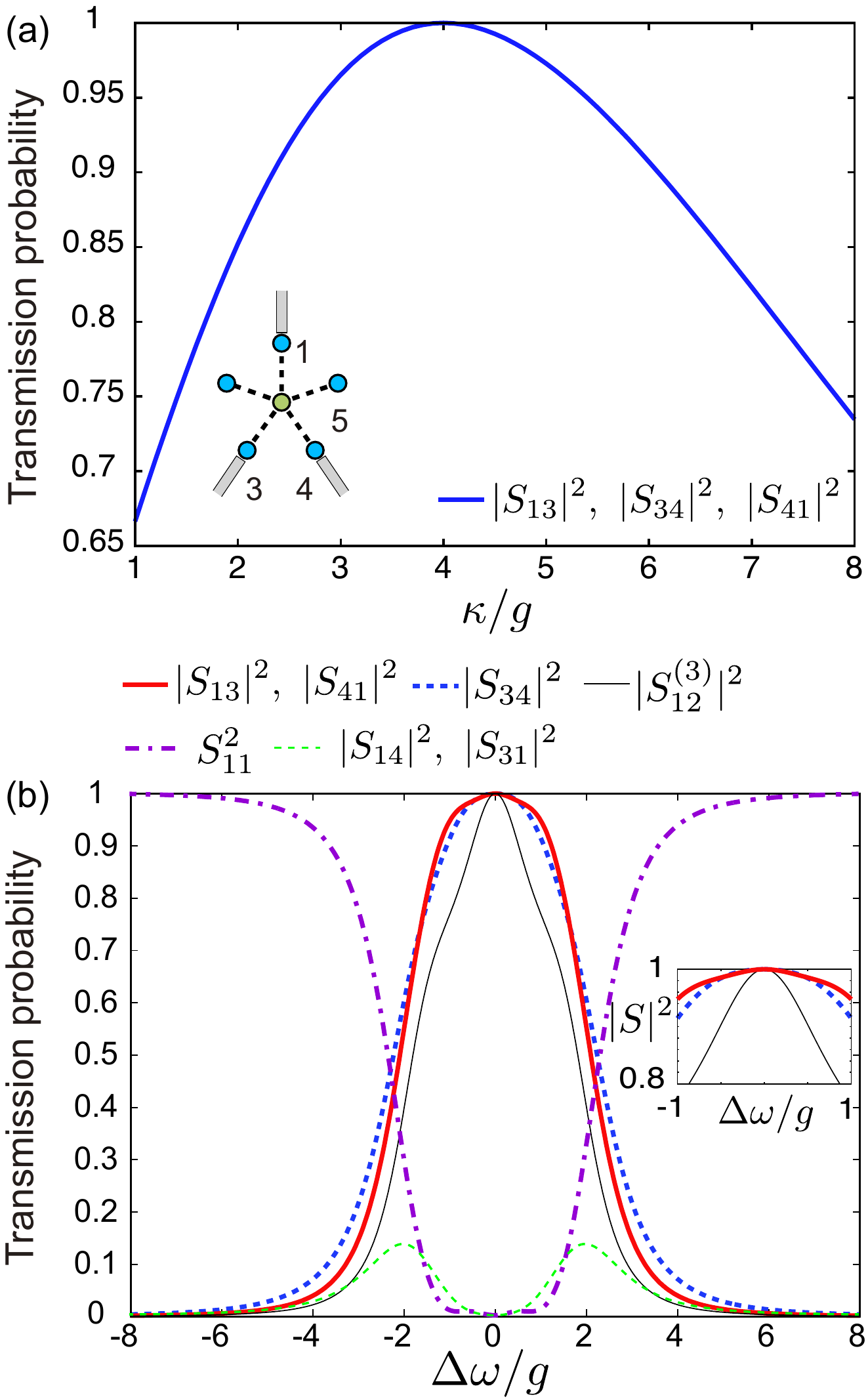}
\end{center}
\caption{
Circulation properties for the case of $N=5$, $\kappa_{1,3,4}=\kappa$, and $\kappa_{2,5}=0$.
(a) Dependence of the forward transmission probabilities ($|S_{13}|^2,\ |S_{34}|^2$ and $|S_{41}|^2$) on $\kappa$ for $\Delta\omega=0$.
The inset shows the system configuration.
(b) Dependence of the forward and backward transmission and reflection probabilities on $\Delta \omega$ for $\kappa=4g$.
The thin black line represents $|S_{12}|^2$ for $N=3$, denoted by $|S_{12}^{(3)}|^2$.
The inset is a closeup around $\Delta \omega/g=0$.
}
\label{S2_omega_kappa_5site_5_7_17}
\end{figure}

Figure~\ref{S2_omega_kappa_5site_5_7_17}(a) shows the dependence of the forward transmission probabilities on $\kappa$ for $N=5$ without detuning, $\Delta\omega=0$.
The transmission lines are coupled to resonators~1, 3 and 4.
We numerically confirm that $|S_{13}|=|S_{34}|=|S_{41}|$.
These equalities are analytically derived in Appendix~\ref{Equalities in S-matrix}.
The forward transmission probabilities become unity at $\kappa=4g$.
This value of $\kappa$ is twice larger than the ideal value of $\kappa (=2g)$ for $N=3$.
Figure~\ref{S2_omega_kappa_5site_5_7_17}(b) shows the dependence of the forward and backward transmission and reflection probabilities on detuning $\Delta \omega$ with $\kappa=4g$. 
It is seen that the forward transmission probabilities for $N=5$ are higher than that for $N=3$.
$S_{33}$ and $S_{44}$ ($S_{43}$) exhibit similar $\Delta \omega$-dependence to $S_{11}$ ($S_{14}$ and  $S_{31}$), and they are not shown here.

\begin{figure}[h!]
\begin{center}
\includegraphics[width=7.7cm]{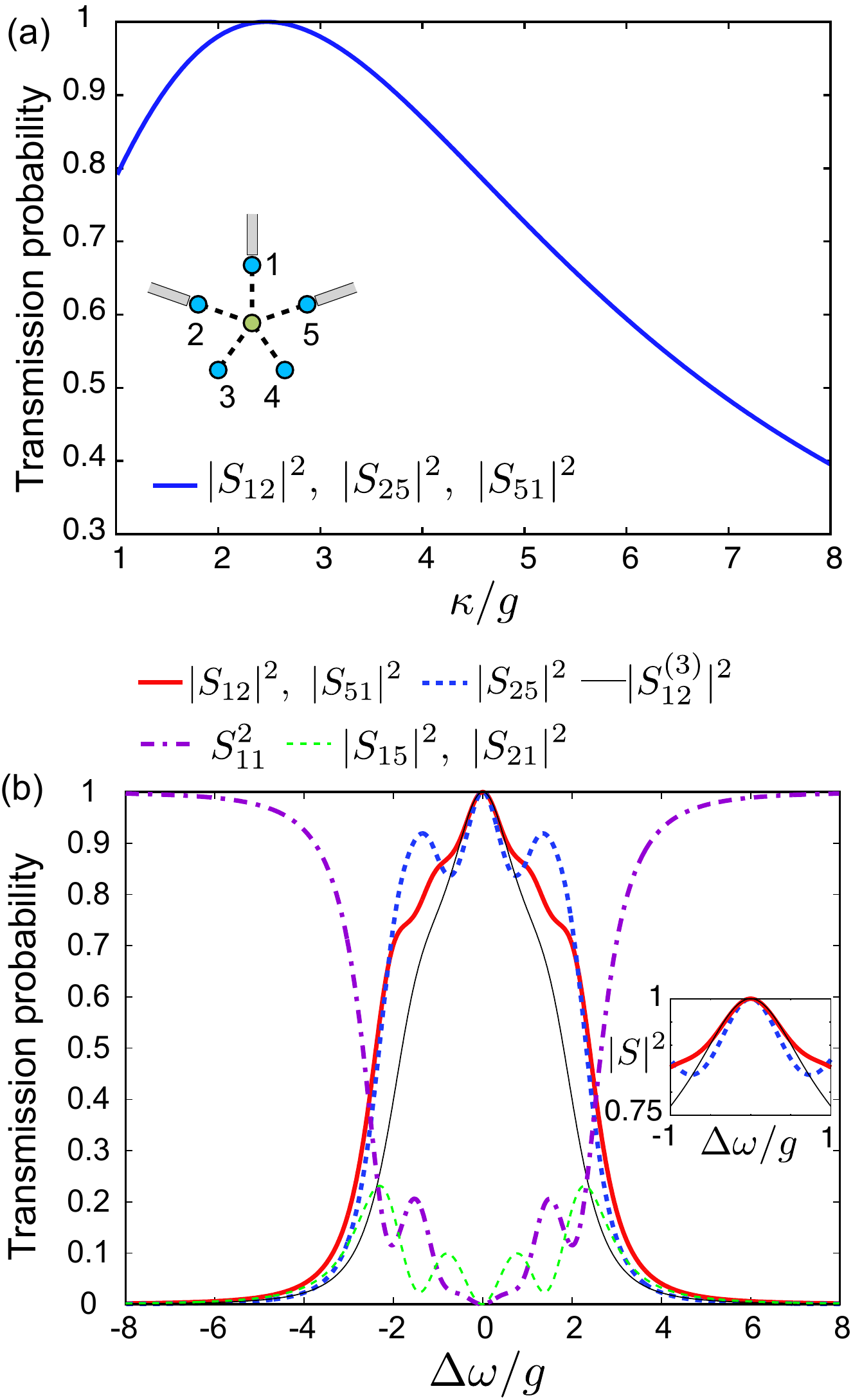}
\end{center}
\caption{
Circulation properties for the case of $N=5$, $\kappa_{1,2,5}=\kappa$, and $\kappa_{3,4}=0$.
(a) Dependence of the forward transmission probabilities (~$|S_{12}|^2,\ |S_{25}|^2$ and $|S_{51}|^2$ ) on $\kappa$ for $\Delta\omega=0$. 
The inset shows the system configuration.
(b) Dependence of the forward and backward transmission and reflection probabilities on $\Delta \omega$ for $\kappa=2.472g$.
The black line represents $|S_{12}^{(3)}|^2$.
The inset is a closeup around $\Delta \omega/g=0$. 
}
\label{S2_omega_kappa_5site_2_5_7_17}
\end{figure}

The robustness of the routing efficiency against detuning depends on which resonators the transmission lines are attached to.
In Fig.~\ref{S2_omega_kappa_5site_2_5_7_17}, we make the same plot as Fig.~\ref{S2_omega_kappa_5site_5_7_17} except that the transmission lines are attached to resonators~1,2 and 5.
Figure~\ref{S2_omega_kappa_5site_2_5_7_17}(a) shows the dependence of
the forward transmission probabilities on $\kappa$.
The optimal value of $\kappa$, with which $|S_{12}|$, $|S_{25}|$, and $|S_{51}|$ become unity, is approximately $2.472 g$.
Figure~\ref{S2_omega_kappa_5site_2_5_7_17}(b) shows the dependence of the forward and backward transmission and the reflection probabilities on detuning for $\kappa=2.472g$. 
The transmission probabilities are comparable to that for $N=3$ for $|\Delta\omega/g|<0.5$ as seen in the inset of Fig.~\ref{S2_omega_kappa_5site_2_5_7_17}(b).
Therefore, the configuration in Fig.~\ref{S2_omega_kappa_5site_5_7_17}(a) is more desirable than the one in Fig.~\ref{S2_omega_kappa_5site_2_5_7_17}(a).
$S_{22}$ and $S_{55}$  ($|S_{52}|$) exhibit a similar $\Delta \omega$-dependence to $S_{11}$ ($|S_{15}|$ and $|S_{21}|$), and they are not shown here.
The results for $N=7$ system is shown in Appendix~\ref{Results for N7}.

\subsection{four-resonator system}
\begin{figure}[h!]
\begin{center}
\includegraphics[width=8cm]{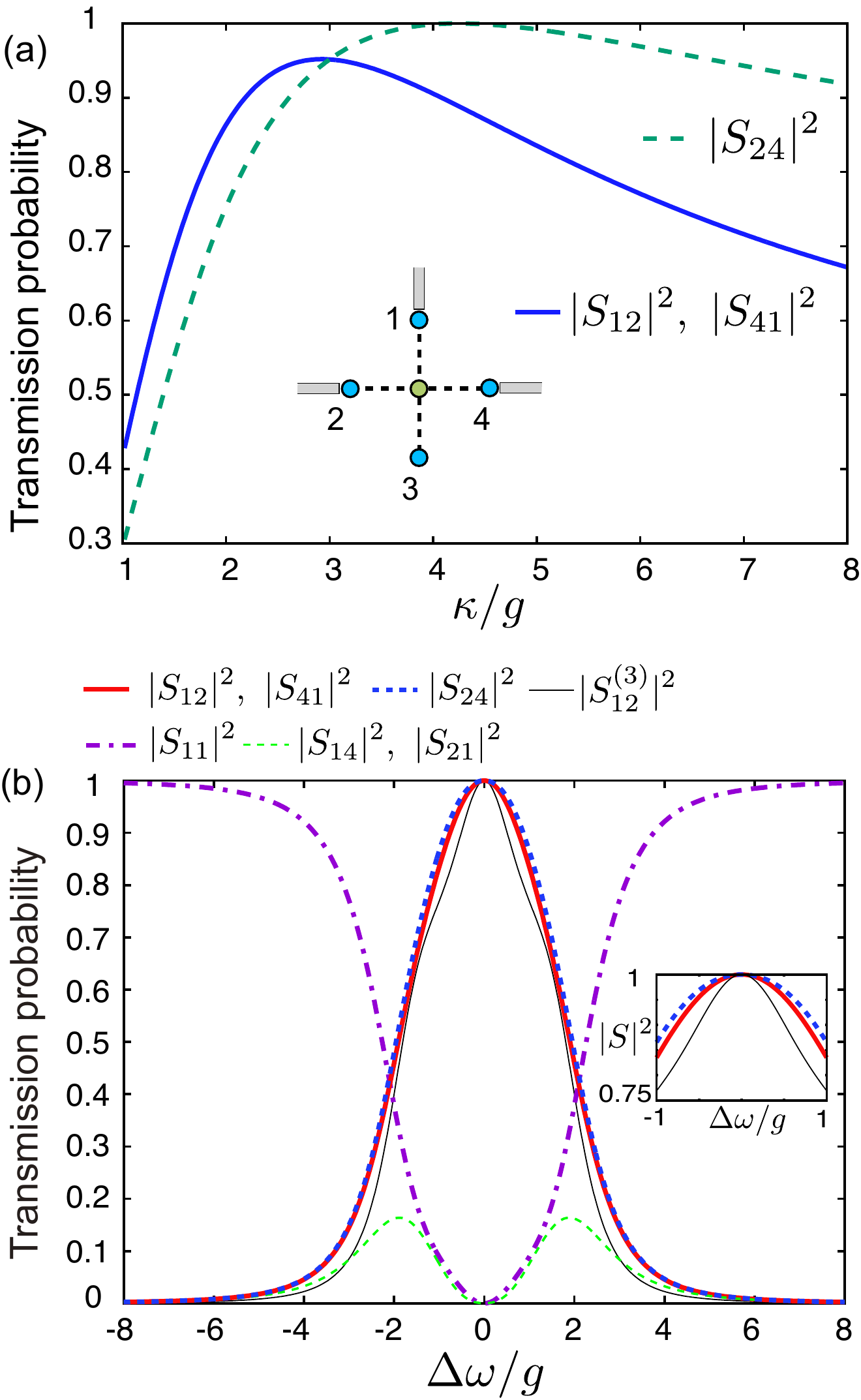}
\end{center}
\caption{
Circulation properties for the case of $N=4$ and $\kappa_{3}=0$.
(a) Dependence of the forward transmission probabilities ( $|S_{12}|^2,\ |S_{24}|^2$ and $|S_{41}|^2$ ) on $\kappa$ for $\Delta\omega=0$ when $\kappa_{1,2,4}=\kappa$.
The inset shows the system configuration.
(b)~Dependence of the forward and backward transmission and reflection probabilities on $\Delta \omega$ for the case of $\kappa_1=2.14g$ and $\kappa_{2,4}=4.24g$.
}
\label{S_omega_kappa_4_6site_6_10_17}
\end{figure}

Figure~\ref{S_omega_kappa_4_6site_6_10_17}(a) shows the dependence of  the forward transmission probabilities on $\kappa$ for $N=4$ without detuning.
The transmission lines are coupled to resonators 1, 2 and 4.
Under the condition of $\kappa_1= \kappa_2= \kappa_4$, it is seen that 
$S_{12}$, $S_{24}$, $S_{41}$ do not reach unity simultaneously.
Therefore, this system does not work as a circulator.

Now we consider the case with $\kappa_1\ne\kappa_2=\kappa_4$.
We optimize $\kappa_1$ and $\kappa_2$ so that  the product of the forward transmission amplitudes defined by $|S_{12}S_{24}S_{41}|$ is maximized.
Figure~\ref{S_omega_kappa_4_6site_6_10_17}(b) shows the dependence of the forward and backward transmission and reflection probabilities on the detuning for $\kappa_1=2.14g$ and $\kappa_{2,4}=4.24g$.
It is seen that the forward transmission probabilities are almost unity for $\Delta \omega=0$, and the routing efficiency is robust against the detuning more than the circulator for $N=3$.
$S_{22}$ and $S_{44}$ ($|S_{42}|$) show a similar $\Delta \omega$-dependence to $S_{11}$ ($|S_{14}|$ and $|S_{21}|$), and they are not shown here.

\subsection{six-resonator system}
\label{six-resonator system}
\begin{figure}[h!]
\begin{center}
\includegraphics[width=8cm]{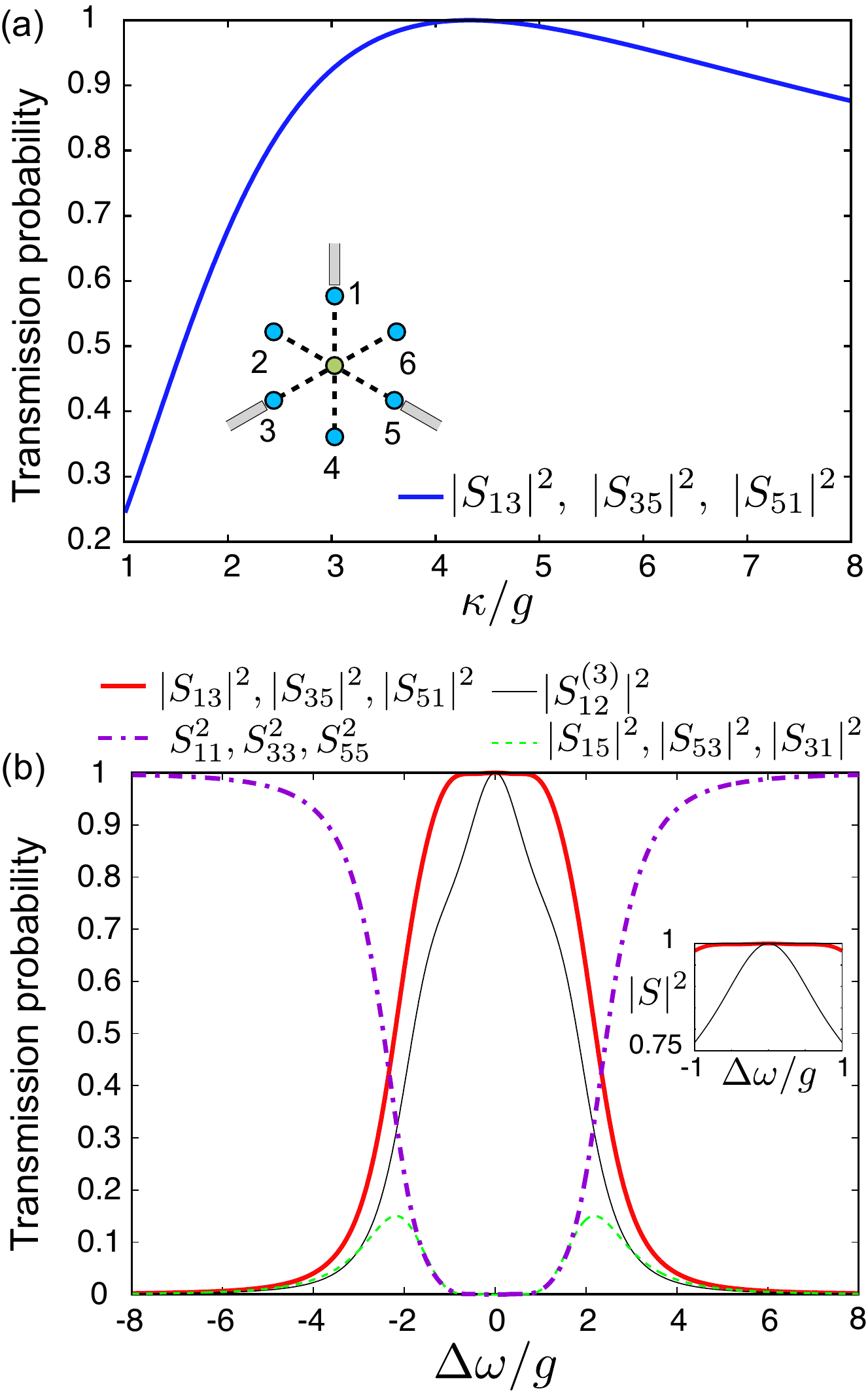}
\end{center}
\caption{
Circulation properties for the case of $N=6$, $\kappa_{1,3,5}=\kappa$, and $\kappa_{4,6}=0$.
(a) Dependence of the forward transmission probabilities (~$|S_{13}|^2,\ |S_{35}|^2$ and $|S_{51}|^2$ ) on $\kappa$ for $\Delta\omega=0$.
The inset shows the system configuration.
(b) Dependence of the forward and backward transmission and reflection probabilities on $\Delta \omega$ for $\kappa=4.328g$.
The black line represents $|S_{12}^{(3)}|^2$.
The inset is a closeup around $\Delta \omega/g=0$. 
}
\label{S2_omega_kappa_6site_2_25_18}
\end{figure}

Figure~\ref{S2_omega_kappa_6site_2_25_18}(a) shows the dependence of the transmission probabilities on $\kappa$ for $N=6$ without detuning.
The transmission lines are coupled to resonators 1, 3 and 5.
As expected from the rotational symmetry,  $|S_{13}|=|S_{35}|=|S_{51}|$.
The transmission probabilities become unity when $\kappa \simeq 4.328g$.

Figure~\ref{S2_omega_kappa_6site_2_25_18}(b) shows the dependence of the transmission and reflection probabilities on the detuning for $\kappa=4.328g$. 
As seen in the inset, the routing efficiency is obviously robust against the detuning compared to the systems with 
other cavity number $N$.

\subsection{Systems with large number of resonators}
\begin{figure}[h!]
\begin{center}
\includegraphics[width=8cm]{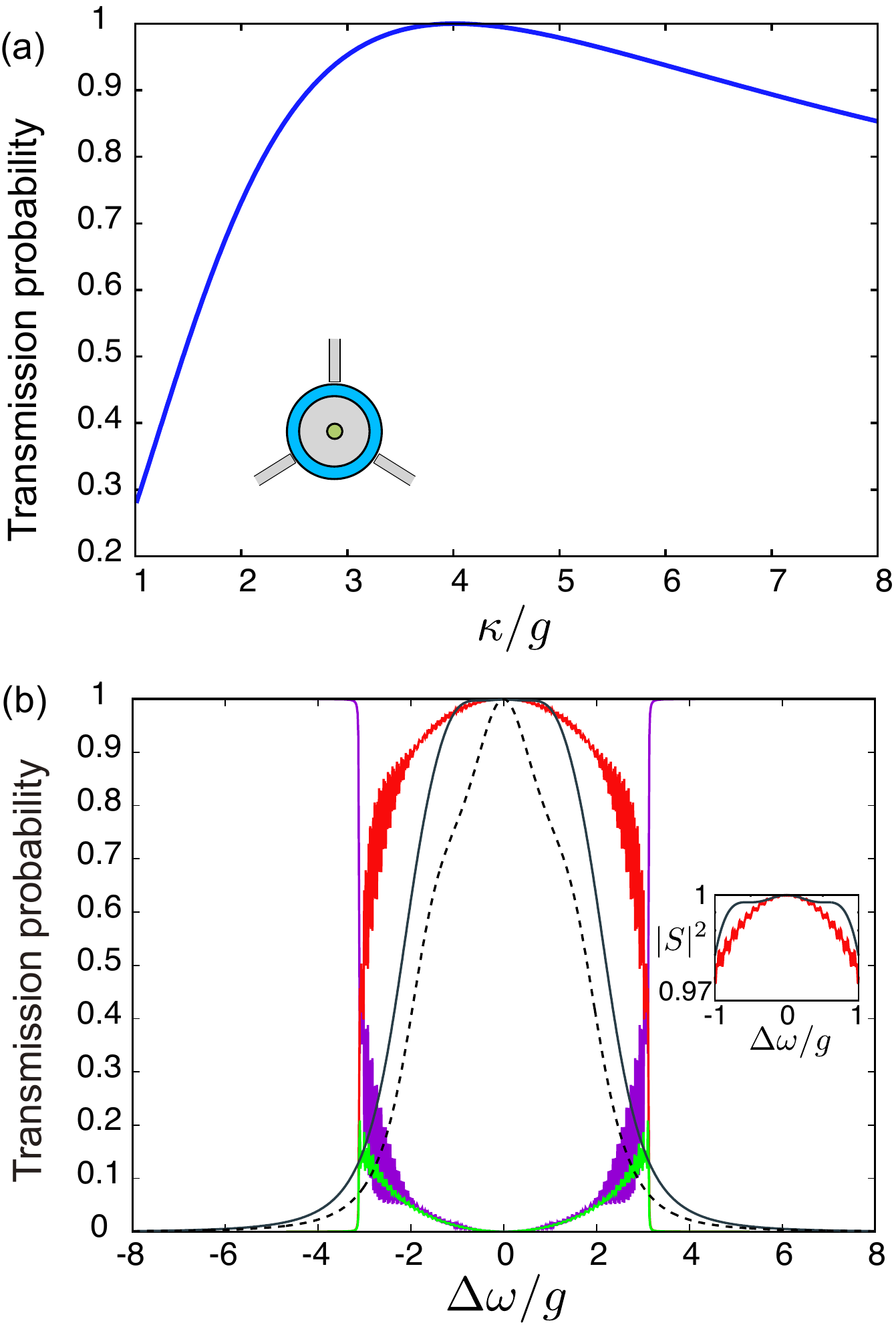}
\end{center}
\caption{
Circulation properties for the case of $N=195$, $\kappa_{1,66,131}=\kappa$, and $\kappa_{i}=0$ for $i\ne1,66,131$.
(a) Dependence of the forward transmission probabilities on $\kappa$ for $\Delta\omega=0$.
The inset schematically shows the system configuration for large $N$.
(b)~Dependence of the forward and backward transmission and reflection probabilities on $\Delta \omega$ for $\kappa\simeq 4g$.
The red line is for the forward transmission probability.
The green and the purple lines are for the backward transmission and reflection probabilities, respectively.
The black solid and dashed lines are for the forward transmission probabilities for $N=6$ system in Fig.~\ref{S2_omega_kappa_6site_2_25_18}(b)
and $N=3$, respectively.
The right inset is the closeup around $\Delta \omega/g=0$.
}
\label{S2_omega_kappa_large_site_2_25_18}
\end{figure}

To see an asymptotic property we examine large $N$ systems.
The transmission lines are attached in a way that the system is geometrically symmetric against $2\pi/3$ rotation for concreteness.
Figure~\ref{S2_omega_kappa_large_site_2_25_18}(a) shows the dependence of the transmission probabilities on $\kappa$ for $N=195$ without detuning.
The transmission probabilities become unity when $\kappa \simeq 4g$.
The dependence of the transmission probabilities on $\Delta \omega$ asymptotically changes to the ones shown in Fig.~\ref{S2_omega_kappa_large_site_2_25_18}(b) when $N$ becomes sufficiently large. These asymptotic profiles do not depend much on the parity of $N$ and where the transmission lines are attached as long as they are separated from each other sufficiently. 
The transmission and the reflection probabilities for $N=192$ show the $\Delta \omega$-dependence qualitatively similar to the ones for $N=195$, although they are not exhibited here.
Note that the system with $N=6$ is more robust against the detuning around $\Delta \omega/g=0$ than these large $N$ systems.

The forward transmission probability for large $N$ system is close to unity for $-\pi g<\Delta \omega<\pi g$. The asymptotic value of the bandwidth of $2\pi g$ is attributed to the energy  band of the GR cluster, which is from $-\hbar g \pi$ to $\hbar g \pi$. The incoming microwave can enter to the GR cluster and can propagate as a plane wave if its energy is in that range, otherwise it is reflected. 

\subsection{Effects of parameter fluctuations}
\begin{figure}[h!]
\begin{center}
\includegraphics[width=6.5cm]{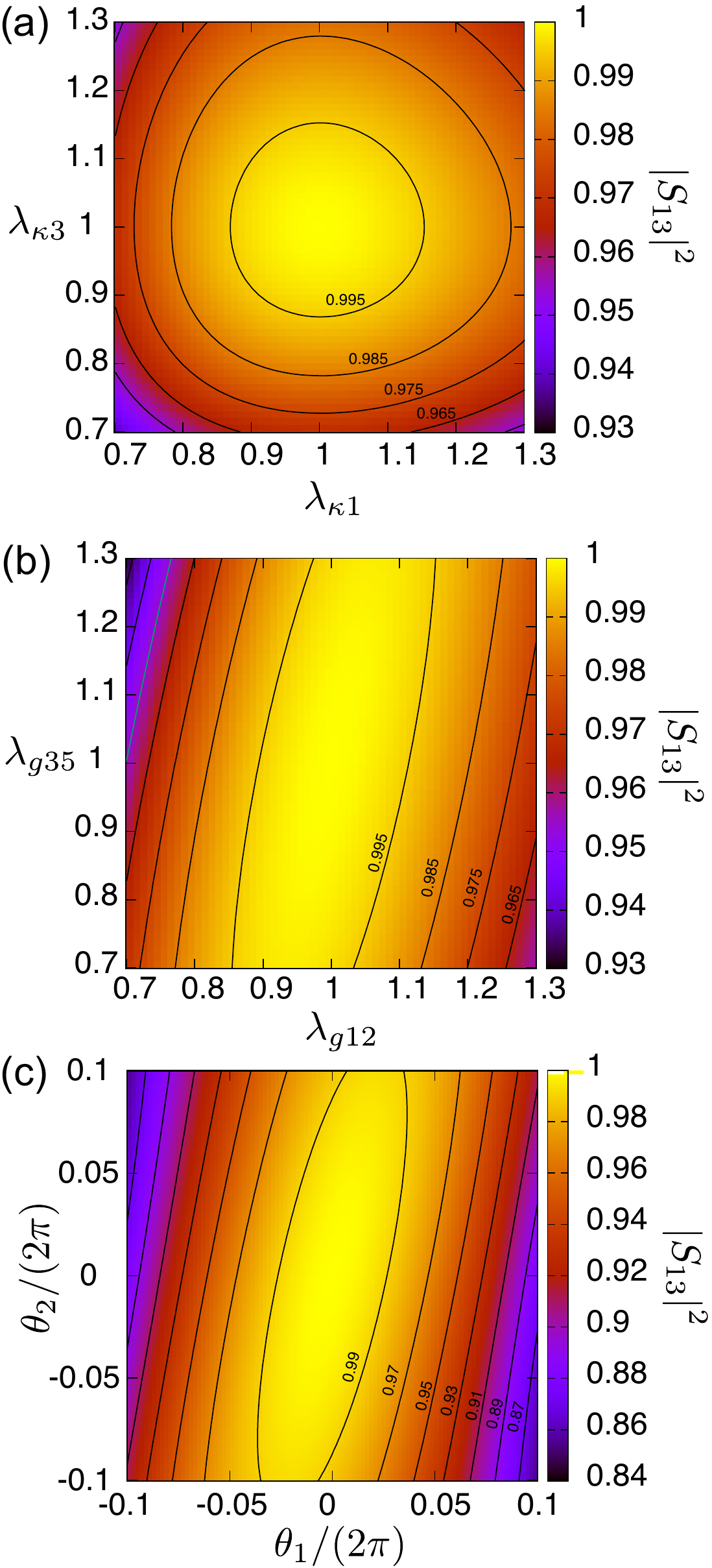}
\end{center}
\caption{
Effects of the fluctuations in the system parameters. The system with $N=6$ in Fig.~\ref{S_omega_kappa_4_6site_6_10_17}(c) is investigated.
Dependence of $|S_{13}|^2$ on (a) $\lambda_{\kappa1}$ and $\lambda_{\kappa3}$,
(b) $\lambda_{g13}$ and $\lambda_{g35}$,
(c) $\theta_{1}$ and $\theta_{2}$.
The values next to the contour lines indicate the values of $|S_{13}|^2$.
}
\label{S_TI_stability_7_6_17}
\end{figure}

Here, we observe the effects of fluctuations of the system parameters. To observe the effects of fluctuation in $\kappa$, we replace  $\kappa_1$ and $\kappa_3$ with $\lambda_{\kappa1}\kappa_1$ and $\lambda_{\kappa3}\kappa_3$, respectively, for the six-resonator system discussed in Sec.~\ref{six-resonator system}.
Figure~\ref{S_TI_stability_7_6_17}(a) shows the dependence of $|S_{13}|^2$ on $\lambda_{\kappa1}$ and $\lambda_{\kappa3}$.
$|S_{13}|^2$ is approximately 0.93 when $\kappa_1$ and $\kappa_3$ have 30\% of inhomogeneity.
Transmission probabilities $|S_{35}|^2$ and $|S_{51}|^2$ (not shown) are higher than 0.965 in the same range of  $\lambda_{\kappa1}$ and $\lambda_{\kappa3}$.

To observe the effects of the fluctuation in $g$, we make the following replacements:
$\overline{g}_{13/31}\rightarrow\lambda_{g13}\overline{g}_{13/31}$ and $\overline{g}_{35/53}\rightarrow\lambda_{g35}\overline{g}_{35/53}$.
Figure~\ref{S_TI_stability_7_6_17}(b) shows the dependence of $|S_{13}|^2$ on $\lambda_{g13}$ and $\lambda_{g35}$.
$|S_{13}|^2$ is approximately 0.93 when the coupling strengths have 30\% of inhomogeneity.
Transmission probability $|S_{35}|^2$ is higher than 0.99 and $|S_{51}|^2$ is higher than 0.93 in the same range of $\lambda_{g13}$ and $\lambda_{g35}$ (not shown).

Finally, to observe the effects of the fluctuation in the phase of $g$, 
we replace $\overline{g}_{13}$ and $\overline{g}_{35}$ with $e^{i\theta_1}\overline{g}_{13}$, and $e^{i\theta_2}\overline{g}_{35}$.
Figure~\ref{S_TI_stability_7_6_17}(c) shows the dependence of $|S_{13}|^2$ on $\theta_1$ and $\theta_2$. 
Transmission probability $|S_{35}|^2$ is higher than 0.96 and $|S_{51}|=|S_{13}|$ 
in the same range of $\theta_1$ and $\theta_2$, although they are not shown here.
It is seen that $|S_{13}|$ is sensitive to $\theta_1$ compared to $\theta_2$.


\subsection{Properties of circulator based on GR hoppings}
Here we discuss the general properties of the circulator based on GR hopping with transmission lines A, B and~C.
As we observed in Sec.~\ref{five-resonator system}, several equalities hold in the transmission probabilities for odd $N$.
We numerically confirmed that $S_{AB}=S_{BC}=S_{CA}$, $S_{AC}=S_{CB}=S_{BA}$ and $S_{AA}=S_{BB}=S_{CC}$ for $\Delta\omega=0$ irrespectively of the resonators to which the transmission lines are attached. 
The equalities for $N=5$ and 7 are derived analytically in Appendix~\ref{Equalities in S-matrix}.

On the other hand, these equalities do not generally hold for even $N$. However similar equalities hold  when the transmission lines are attached in a way that the system is geometrically symmetric against $2\pi/3$ rotation (see also Appendix~\ref{Equalities in S-matrix} for the $N=6$ case).
Importantly, the equalities which is desirable for a circulator, $|S_{AB}|=|S_{BC}|=|S_{CA}|$, hold for the six-resonator system as seen in Fig.~\ref{S_omega_kappa_4_6site_6_10_17}(b).

\subsection{Drawback and other possible configurations}
The drawback of our scheme is the number of the required couplings between resonators, $N(N-1)/2$, increasing with the number $N$ of resonators.
Unwanted crosstalk between resonators and unwanted excitations of the coupler should be avoided.
Thus, it becomes experimentally more challenging when $N$ increases.
We discuss the physical realization of our circulator with a concrete circuit model in Appendix~\ref{Effective photon Hamiltonian}.

In the present article we have mainly studied the systems with small number of resonators~$N$ and with only three transmission lines. 
Only a part of the possible arrangements of the transmission lines have been examined, although there are many other choices in the arrangement for cases with large $N$.
The optimized arrangement will be studied for larger~$N$ in an experimentally feasible range elsewhere, and routing microwaves based on other types of long range hopping will be also studied.

\section{conclusion}
\label{conclusion and discussion}
We have proposed an on-chip microwave circulator based on the Gebhard-Ruckenstein (GR) hopping.
The linear energy dispersion of the GR cluster gives rise to a chiral propagation of a microwave, and thus can work as a circulator when transmission lines are attached.
Our circulators composed of more than three resonators can have a wider operating bandwidth than that composed of three resonators.
Especially, the circulator composed of six resonators with the three fold rotational symmetry has a remarkably wide operating bandwidth. 
The robustness of the routing efficiency against the inhomogeneity in the system parameters has also been examined.

\section*{Acknowledgments}
We acknowledge the support from JST ERATO (Grant No. JPMJER1601).
KK is grateful to JSPS for a support from JSPS KAKENHI (Grant No. 16K05497).

\appendix

\section{Derivation of $S$-matrix}
\label{Smatrix}
We derive the $S$-matrix of the system, 
which is described by the Hamiltonian of Eq.~(\ref{H_6_6_17})~[\citenum{Koshino2012},\citenum{Koshino2013}].
The Heisenberg equation of motion of $b_{m,k}$ is represented as
\begin{eqnarray}
\frac{d}{dt}b_{m,k}(t) = -i v k b_{m,k}(t) - i\sqrt{\frac{v \kappa_m}{2\pi}} a_m(t).
\label{Hei_6_23_17}
\end{eqnarray}
We define the real-space representation $\widetilde{b}_{m,r}$ of the transmission line field as
\begin{eqnarray}
\widetilde{b}_{m,r}=\frac{1}{\sqrt{2\pi}}\int_{-\infty}^\infty dk e^{ikr} b_{m,k}.
\end{eqnarray}
Tildes are used to distinguish the operator in $k$-space representation from the one in $r$-space representation in this section.
In this representation, the field interacts with
resonator at $r=0$, and the $r<0$ ($r>0$) region corresponds to the incoming (outgoing) field. 
The input and output field operators of the transmission line field are defined by
\begin{eqnarray}
\widetilde{b}_{m,r}^{(\rm in)}(t) = \widetilde{b}_{m,-r}(t),\nonumber\\
\widetilde{b}_{m,r}^{(\rm our)}(t) = \widetilde{b}_{m,r}(t),
\label{b_in_out_8_18_17}
\end{eqnarray}
respectively, where $r>0$.
Using Eqs.~(\ref{Hei_6_23_17}) and (\ref{b_in_out_8_18_17}), the input-output relation is derived as~\cite{Koshino2012,Koshino2013}
\begin{eqnarray}
\widetilde{b}_{m,+0}^{(\rm out)}(t) = \widetilde{b}_{m,v t}^{(\rm in)}(0) - i\sqrt{\frac{\kappa_m}{v}} a_m(t).
\label{in_out_6_28_17}
\end{eqnarray}
In Eq.~(\ref{in_out_6_28_17}) $\widetilde{b}_{m,v t}^{(\rm in)}(0)=\widetilde{b}_{m,+0}^{(\rm in)}(t)$.
The Heisenberg equation of motion of $a_m$ is represented as
\begin{eqnarray}
\frac{d}{dt}a_m &=& \Big{(} -i\omega_m - \frac{\kappa_m}{2} \Big{)} a_m
- i \sum_{n (\ne m)} g_{m,n}(t) a_n\nonumber\\
&&- i \sqrt{v\kappa_m} \widetilde{b}_{m,vt}^{(\rm in)}.
\label{a_6_6_17}
\end{eqnarray}

So far we discussed the operator equations.
Here we assume that an input microwave is applied from the transmission line attached to the $p$-th resonator.
We consider a continuous mode version of a coherent state:
\begin{eqnarray}
|\Psi_i\rangle = \mathcal{N}\exp\Big{[}\int_{0}^{\infty}dr E_{\rm in}(r) \big{(}\widetilde{b}_{p,r}^{(\rm in)}\big{)}^\dagger \Big{]} |0\rangle,
\label{Psii_6_28_17}
\end{eqnarray}
with the overall vacuum state $|0\rangle$ and a normalization constant $\mathcal{N}$.
Considering that the input wave propagates in the negative $r$ direction,
$E_{\rm in}(r)$ represents the input microwave at the initial moment as given by
\begin{eqnarray}
E_{\rm in}(r) = E e^{- i \omega_{\rm in} r/v},
\end{eqnarray}
where $E$ and $\omega_{\rm in}$ are the amplitude and the angular frequency of the incident microwave, respectively.
We assume that at the initial moment the resonators and the transmission lines except for the input one are unexcited, and the input microwave has not arrived at resonator~$p$ yet.
$|\Psi_i\rangle$ is in a coherent state and therefore an eigenstate of the initial field operator $\widetilde{b}_{p,r}(0)$.
It is confirmed that 
\begin{eqnarray}
\widetilde{b}_{m,vt}^{({\rm in})}|\Psi_i\rangle = E_{\rm in}(vt) \delta_{m,p} |\Psi_i\rangle 
= E e^{-i\omega_{\rm in}t} \delta_{m,p} |\Psi_i\rangle\nonumber\\
\label{blvt_6_28_17}
\end{eqnarray}
using Eqs.~(\ref{b_in_out_8_18_17}) and (\ref{Psii_6_28_17}).

We rewrite Eq.~(\ref{a_6_6_17}) as 
\begin{eqnarray}
\frac{d}{dt}A_m &=& \Big{(} i\Delta \omega  - \frac{\kappa_m}{2} \Big{)} A_m
- i \sum_{n (\ne m)} g_{m,n}(t) 
e^{i(\omega_m-\omega_n)t} A_n \nonumber\\
&&
- i \sqrt{v\kappa_m} \widetilde{b}_{m,vt}^{({\rm in})}  e^{i(\omega_m + \Delta \omega)t}
\label{Al_6_6_17}
\end{eqnarray}
with $A_m$ defined by
\begin{eqnarray}
A_m(t) = e^{i(\omega_m + \Delta \omega)t} a_m(t),
\end{eqnarray}
where $\Delta \omega$ is the detuning of the incident microwave with angular frequency $\omega_{\rm in}$, that is, $\omega_{\rm in}=\omega_p + \Delta \omega$.
Using Eqs.~(\ref{g_6_6_17}) and~(\ref{Al_6_6_17}) and the rotating wave approximation, we obtain
\begin{eqnarray}
\frac{d}{dt}A_m &=& \Big{(} i\Delta \omega  - \frac{\kappa_m}{2} \Big{)} A_m - i \sum_{n (\ne m)} \overline{g}_{m,n}
e^{-i\theta_{m,n}} A_n\nonumber\\
&&- i \sqrt{v\kappa_m} \widetilde{b}_{m,vt}^{({\rm in})}e^{i(\omega_m + \Delta \omega)t}.
\label{Al_6_6_17_2}
\end{eqnarray}

Taking the expectation value of Eq.~(\ref{a_6_6_17}) with respect to $|\Psi_i\rangle$ we obtain
\begin{eqnarray}
\frac{d}{dt}\langle A_m\rangle &=& \Big{(} i\Delta \omega  - \frac{\kappa_m}{2} \Big{)} \langle A_m\rangle - i \sum_{n (\ne m)} \overline{g}_{m,n}
e^{-i\theta_{m,n}} \langle A_n\rangle\nonumber\\
&&- i \sqrt{v\kappa_m} E \delta_{m,p}.
\label{Al_6_28_17}
\end{eqnarray}
To obtain the stationary solution in the rotating frame we put $d\langle A_m\rangle/dt=0$.
Then Eq.~(\ref{Al_6_28_17}) is rewritten in the matrix form as 
\begin{eqnarray}
\mathcal{G} \overrightarrow{\langle{A}\rangle} = -i \sqrt{v\kappa_p} E  \overrightarrow{\phi_p},
\label{Goriginal_7_6_17}
\end{eqnarray}
where the ($m,n$) element of matrix $\mathcal{G}$ is given by
\begin{eqnarray}
\mathcal{G}_{m,n}= \begin{cases}
\kappa_m/2 - i\Delta \omega & (n=m) \\
i \overline{g}_{m,n} e^{-i\theta_{m,n}} & (n\ne m),
\end{cases}
\end{eqnarray}
and
\begin{eqnarray}
\overrightarrow{\langle A\rangle} = \big{(} \langle A_1\rangle, \cdots, \langle A_N\rangle \big{)},
\end{eqnarray}
and the $p$-th component of $\overrightarrow{\phi_p}$ is 1, while the others are 0.
Then $\overrightarrow{\langle A\rangle}$ is written with the inverse of matrix $\mathcal{G}$ as
\begin{eqnarray}
\overrightarrow{\langle A\rangle} = -i \sqrt{v\kappa_p} E \mathcal{G}^{-1} \overrightarrow{\phi}_p.
\label{Al_6_6_17_4}
\end{eqnarray}

We multiply Eq.~(\ref{in_out_6_28_17}) by $e^{i(\omega_m+\Delta\omega)t}$ and take the expectation value with respect to $|\Psi_i\rangle$ to obtain
\begin{eqnarray}
\langle \widetilde{B}_{m,+0}^{(\rm out)}(t) \rangle = \langle \widetilde{B}_{m,vt}^{(\rm in)} \rangle 
- i \sqrt\frac{\kappa_m}{v} \langle A_m(t) \rangle,
\label{in_out_6_28_17_2}
\end{eqnarray}
where 
\begin{eqnarray}
\widetilde{B}_{m,+0}^{(\rm out)}(t) &=& \widetilde{b}_{m,+0}^{(\rm out)}(t) e^{i(\omega_m+\Delta\omega)t}\nonumber\\
\widetilde{B}_{m,vt}^{(\rm in)} &=& \widetilde{b}_{m,vt}^{(\rm in)} e^{i(\omega_m+\Delta\omega)t}.
\label{B_2_27_18_2}
\end{eqnarray}
Using the stationary solution of $\langle A_m \rangle$ given by Eq.~(\ref{Al_6_6_17_4}), the stationary solution of 
$\langle \widetilde{B}_{m,+0}^{(\rm out)} \rangle$ is given by 
\begin{eqnarray}
\langle \widetilde{B}_{m,+0}^{(\rm out)} \rangle = \langle \widetilde{B}_{m,vt}^{(\rm in)} \rangle 
- i \sqrt{\frac{\kappa_m}{v}}\langle A_m \rangle,
\label{B_2_27_18}
\end{eqnarray}
where 
\begin{eqnarray}
 \langle \widetilde{B}_{m,vt}^{(\rm in)} \rangle = \delta_{m,p} E,
 \label{B_2_27_18_3}
\end{eqnarray}
because of Eq.~(\ref{blvt_6_28_17}).
The elements of $S$-matrix are defined by
\begin{eqnarray}
S_{p,m} = \frac{\langle \widetilde{b}_{m,+0}(t) \rangle}{\langle \widetilde{b}_{p,-0}(t)\rangle}.
\label{S_6_28_17}
\end{eqnarray}
With the use of Eqs.~(\ref{Al_6_6_17_4}), (\ref{B_2_27_18_2}), (\ref{B_2_27_18}), (\ref{B_2_27_18_3}), (\ref{S_6_28_17}) and $\widetilde{b}_{m,v t}^{(\rm in)}(0)=\widetilde{b}_{m,+0}^{(\rm in)}(t)$, the $S$-matrix element is given as
\begin{eqnarray}
S_{p,m} &=& \delta_{p,m}   - \sqrt{\kappa_p\kappa_m} [\mathcal{G}^{-1} \overrightarrow{\phi_p}]_m,\nonumber\\
&=& \delta_{p,m}   - \sqrt{\kappa_p\kappa_m} [\mathcal{G}^{-1}]_{m,p},
\end{eqnarray}
where $[\mathcal{G}^{-1}\overrightarrow{\phi_p}]_m$ denotes the $m$-th component of vector, $\mathcal{G}^{-1}\overrightarrow{\phi_p}$.
This is identical to $[\mathcal{G}^{-1}]_{m,p}$, since the $p$-th component of $\overrightarrow{\phi_p}$ is 1 and the others are 0.

\section{Equalities between $S$-matrix elements}
\label{Equalities in S-matrix} 
We analytically show the equalities between the $S$-matrix elements, which are numerically confirmed in Secs.~\ref{five-resonator system} and \ref{six-resonator system}.
We consider the matrix $\mathcal{G}$ in Eq.~(\ref{Goriginal_7_6_17}) for the system with $N=5$ depicted in Fig.~ \ref{model_6_6_17}.
The matrix $\mathcal{G}$ has the form as
\begin{eqnarray}
\mathcal{G} = \left( \begin{array}{ccccc}
x & a & b & -b & -a  \\
-a & 0 & a & b & -b  \\
-b & -a & x & a & b  \\
b & -b & -a & x & a  \\
a & b & -b & -a & 0  \\
\end{array}
\right),
\label{G_7_6_17}
\end{eqnarray}
with real constants $x$, $a$ and $b$ when $\Delta\omega=0 $.
The matrix elements of $\mathcal{G}^{-1}$ are represented as
\begin{eqnarray}
\big{[}\mathcal{G}^{-1}\big{]}_{31/43/14} &=& \frac{A_5 + B_5 x}{3A_5x + b^2x^3},\nonumber\\
\big{[}\mathcal{G}^{-1}\big{]}_{41/34/13} &=& \frac{A_5 - B_5 x}{3A_5x + b^2 x^3},\nonumber\\
\big{[}\mathcal{G}^{-1}\big{]}_{11/33/44} &=& \frac{A_5 +b^2 x^2}{3A_5x + b^2x^3}
\end{eqnarray}
with 
\begin{eqnarray}
A_5 &=& a^4 - 2a^3b - a^2b^2 + 2ab^3 + b^4, \nonumber\\
B_5 &=& -a^2b + ab^2 + b^3.
\end{eqnarray}
From Eq.~(\ref{S_7_1_17}), we obtain the following equalities between the $S$-matrix elements:
\begin{eqnarray}
S_{13} = S_{34} = S_{41},\nonumber\\
S_{14} = S_{43} = S_{31},\nonumber\\
S_{11} = S_{33} = S_{44}.
\end{eqnarray}
it is confirmed analytically that the same equalities hold in the system with $N=7$.

Now we consider the case with $N=6$ depicted in Fig.~\ref{S_omega_kappa_4_6site_6_10_17}(b).  
Because matrix $\mathcal{G}$ has the following form:
\begin{eqnarray}
\mathcal{G} = \left( \begin{array}{cccccc}
x & a & b & c & b & a  \\
-a & 0 & a & b & c & b  \\
-b & -a & x & a & b & c  \\
-c & -b & -a & 0 & a & b  \\
-b & -c  & -b & -a & x & a  \\
-a & -b & -c & -b & -a & 0  \\
\end{array}
\right),
\label{G_N6_7_7_17}
\end{eqnarray}
for the case $\Delta \omega=0$,
we have the equalities relevant to $S-$matrix such as 
\begin{eqnarray}
\big{[}\mathcal{G}^{-1}\big{]}_{31/53} &=& -\big{[}\mathcal{G}^{-1}\big{]}_{15},\nonumber \\
\big{[}\mathcal{G}^{-1}\big{]}_{35/13} &=& -\big{[}\mathcal{G}^{-1}\big{]}_{51},\nonumber \\
\big{[}\mathcal{G}^{-1}\big{]}_{11} &=& \big{[}\mathcal{G}^{-1}\big{]}_{33} =  \big{[}\mathcal{G}^{-1}\big{]}_{55}.
\end{eqnarray}
From these equalities, we have 
\begin{eqnarray}
S_{13} = S_{35} = S_{51},\nonumber\\
S_{15} = S_{53} = S_{31},\nonumber\\
S_{11} = S_{33} = S_{55}.
\end{eqnarray}

\section{Results for seven-resonator system}
\label{Results for N7}
\begin{figure}[h!]
\begin{center}
\includegraphics[width=7.5cm]{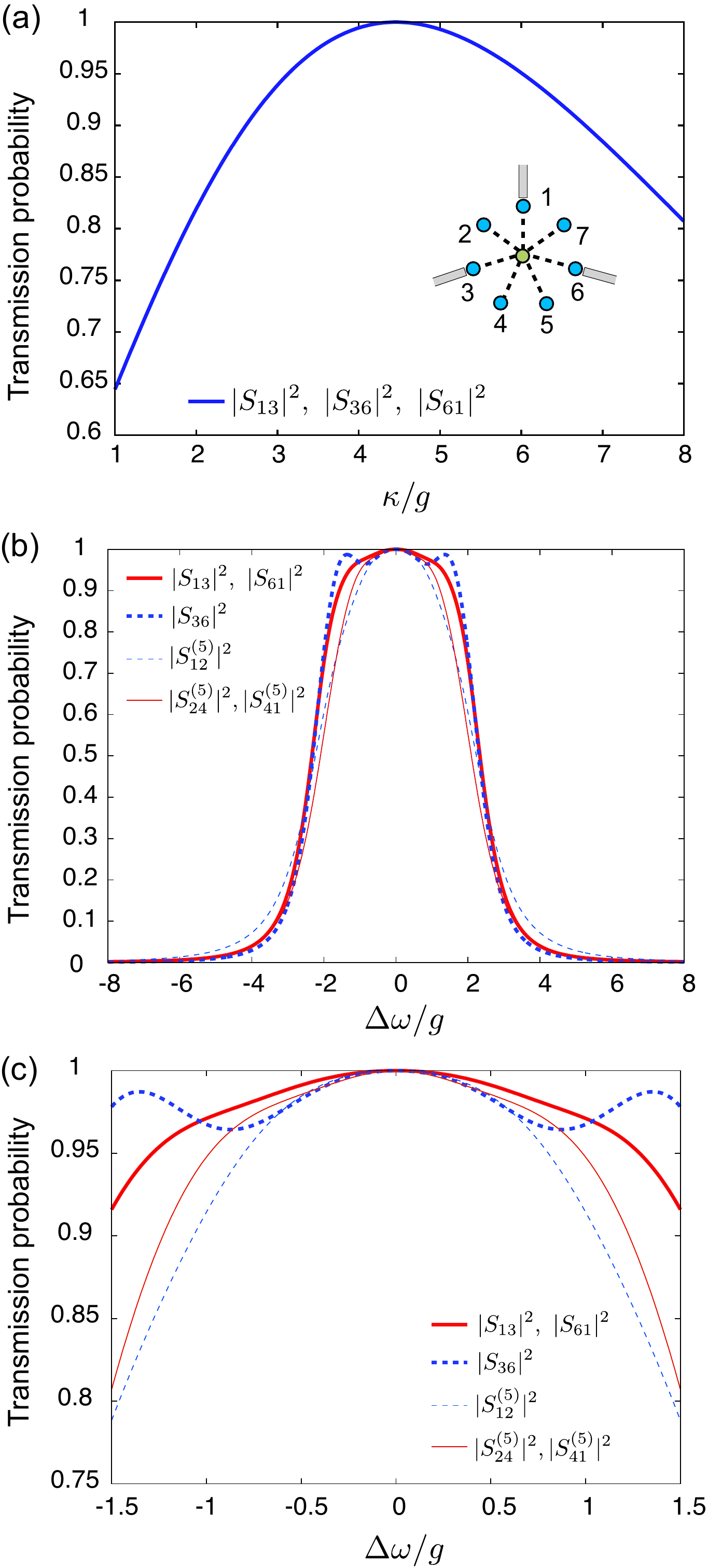}
\end{center}
\caption{
Circulation properties for the case of $N=7$, $\kappa_{1,3,6}=\kappa$, and $\kappa_{2,4,5,7}=0$.
(a)~Dependence of the forward transmission probabilities on the coupling to transmission line $\kappa$ for $\Delta\omega=0$.
The inset shows the system configuration.
(b)~Dependence of the forward transmission probabilities on detuning $\Delta \omega$ for $\kappa=4.45g$.
The thin lines correspond to the system with $N=5$ in Fig.~\ref{S2_omega_kappa_5site_5_7_17}.
(c)~Closeup of (b) around $\Delta \omega/g=0$.
}
\label{S_omega_kappa_site7_5_7_17}
\end{figure}

Figure~\ref{S_omega_kappa_site7_5_7_17}(a) shows the dependence of the forward transmission probabilities on $\kappa$ for $N=7$.
The transmission lines are coupled to resonators 1, 3 and 6 as shown in the inset.
It is numerically confirmed that the forward transmission probabilities, $|S_{13}|^2$, $|S_{36}|^2$ and $|S_{61}|^2$, are identical in spite of the absence of the three fold rotational symmetry and become unity at $\kappa\simeq4.45g$.
We also analytically confirmed the equalities of the forward transmission probabilities in the same manner as
Appendix~\ref{Equalities in S-matrix}.
Figure~\ref{S_omega_kappa_site7_5_7_17}(b) and \ref{S_omega_kappa_site7_5_7_17}(c) show the dependence of the forward transmission probabilities on detuning $\Delta \omega$ for $\kappa=4.45g$. 
The transmission probabilities for $N=7$ are generally higher than those for $N=5$ in Fig.~\ref{S2_omega_kappa_5site_5_7_17}(b) in a wide range of $\Delta \omega$.
The profile of the transmission probabilities for $N=7$ is close to a rectangular shape compared to the one for $N=3$.
The backward transmission and the reflection probabilities for $N=7$ show the $\Delta \omega$-dependence qualitatively similar to the ones for $N=5$ in Fig.~\ref{S2_omega_kappa_5site_5_7_17}(b), although they are not exhibited here.

\section{Physical realization}
\label{Effective photon Hamiltonian}
\begin{figure}[h!]
\begin{center}
\includegraphics[width=5cm]{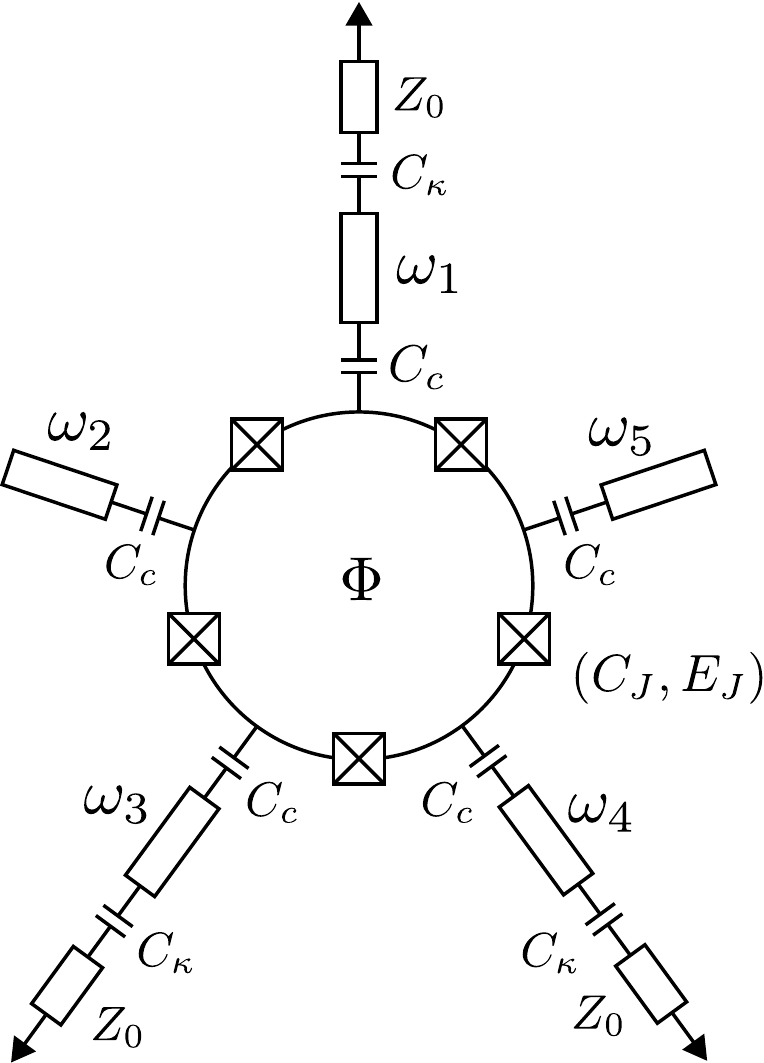}
\end{center}
\caption{
Circuit diagram of a possible physical realization of the system depicted in Fig.\ref{model_6_6_17}.
$Z_0$ is the characteristic impedance of the transmission lines.
$C_\kappa$ is the coupling capacitance between the resonators and the transmission lines.
$C_c$ is the coupling capacitance between the resonators and the Josephson ring.
$C_J$ and $E_J$ are the junction capacitance and the Josephson energy, respectively.
}
\label{model_real_7_7_17}
\end{figure}

In this section we discuss the physical realization of our circulator with a concrete circuit model with $N=5$.
Figure~\ref{model_real_7_7_17} shows a circuit system of a possible physical realization of the circulator.
The system is composed of a Josephson ring, resonators and transmission lines.
The Josephson ring works as the coupler depicted in Fig.~\ref{model_6_6_17}.
A magnetic flux, $\Phi$, is penetrating the Josephson ring.

We derive an effective photon Hamiltonian for the system depicted in Fig.\ref{model_real_7_7_17} in a manner analogous to Ref.\citenum{Koch2010}.
The Hamiltonian of the system is represented by
\begin{eqnarray}
H =  H_{\rm res} + H_{\rm Jring} + H_{\rm int},
\end{eqnarray}
where $H_{\rm res}$, $H_{\rm Jring}$ and $H_{\rm int}$ respectively describe the resonators, the Josephson ring and their interaction.
$H_{\rm res}$ is given by
\begin{eqnarray}
H_{{\rm res}} = \sum_{m=1}^5 \hbar \omega_m a_{m}^\dagger a_m,
\end{eqnarray}
and $H_{\rm Jring}$ is given by
\begin{eqnarray}
H_{\rm Jring} = \frac{1}{2}{\overrightarrow{Q}}^T \mathcal{C}^{-1} {\overrightarrow{Q}} + V({\overrightarrow{\phi}},\Phi),
\end{eqnarray}
with the charge vector ${\overrightarrow{Q}}^T=(Q_1,Q_2,Q_3,Q_4,Q_5)$ and the flux vector ${\overrightarrow{\phi}}=(\phi_1,\phi_2,\phi_3,\phi_4,\phi_5)$, where $Q_\mu$ and $\phi_\mu$ are the charge and the flux on node $\mu$ of the Josephson ring, respectively, and satisfy the commutation relation, $[\phi_{\mu}, Q_{\nu}]=i\hbar\delta_{\mu,\nu}$.
$\mathcal{C}$ is the capacitance matrix with matrix elements $C_{\mu,\mu\pm1}=-C_J$ and $C_{\mu,\mu}=2C_J + C_c$, where $C_J$ and $C_c$ are the junction capacitance and the coupling capacitance between a resonator and the Josephson ring, respectively.
The inductive energy is represented as
\begin{eqnarray}
V({\overrightarrow{\phi}},\Phi) = -E_J \sum^5_{\mu=1} \cos\Big{[} \frac{2\pi}{\Phi_0}
(\phi_{\mu+1} - \phi_{\mu} - \Phi/5)\Big{]},\nonumber\\
\end{eqnarray}
with the Josephson energy $E_J$, where $\mu$ is modulo 5, that is, $\mu=6$ is identical to $\mu=1$.
The interaction Hamiltonian is given by
\begin{eqnarray}
H_{\rm int} = C_c \sum_m \big{(} {\bf e}_m^T \mathcal{C}^{-1} {\overrightarrow{Q}} \big{)} q_{m}\varphi_m,
\label{Hint_7_20_17}
\end{eqnarray}
where ${\bf e}_m$ is the unit vector of which $m$-th element is unity and the others are zero, and
$q_{m} = \sqrt{\omega_m/2} (a_m + a_{m}^\dagger)$,
and $\varphi_m$ is the amplitude of the mode function at the coupling capacitance $C_c$.
In Eq.~(\ref{Hint_7_20_17}), ${\bf e}_m^T C^{-1} {\overrightarrow{Q}}$ corresponds to the voltage of the superconductive island $m$, and $q_{m}\varphi_m$ is the voltage at the end of resonator~$m$.
$H_{\rm int}$ can be represented also as $H_{\rm int}=C_c \overrightarrow{V}_{\rm ring}\cdot \overrightarrow{V}_{\rm res}$ with $\overrightarrow{V}_{\rm ring}= C^{-1} {\overrightarrow{Q}}$ and $\overrightarrow{V}^T_{\rm res}=(q_{1}\varphi_1,q_{2}\varphi_2,\cdots)$.
We rewrite Eq.~(\ref{Hint_7_20_17}) as 
\begin{eqnarray}
H_{\rm int} = 2e C_c \sum_{m,\mu} \xi_{m,\mu}   V_{\rm rms}^{(m)} n_\mu(a_m + a_{m}^\dagger),
\label{Hint_7_24_17}
\end{eqnarray}
where $n_\mu=Q_\mu/(2e)$, and $\xi_{m,\mu} =[\mathcal{C}^{-1}]_{m,\mu}$, and $V_{\rm rms}^{(m)}= \varphi_m\sqrt{\omega_m/2}$.

Now we derive an effective photon lattice Hamiltonian in the dispersive regime, where the coupling between a resonator and the Josephson ring is sufficiently smaller than the energy difference between photonic and circuit excitations.
We assume that the Josephson ring transfers microwaves via intermediate virtual
excitations and remains in its ground state during the operation\cite{Koch2010}.
The effective Hamiltonian is obtained by the canonical transformation
\begin{eqnarray}
H_{\rm ph} &=& P_0 e^{iS}He^{-iS}P_0\nonumber\\
&=&  P_0(H_{{\rm res}} + H_{{\rm Jring}})P_0 + P_0[iS,H_{\rm int}]P_0/2 + O(H_{\rm int}^3),\nonumber\\
\label{Hph_7_21_17}
\end{eqnarray}
where $P_0$ is the projection operator onto the subspace in which the Josephson ring is in its ground state.
Here, $iS$ is defined by
\begin{eqnarray}
iS = \sum_{\alpha,\alpha'} \frac{\langle \alpha'|H_{\rm int}|\alpha\rangle}{E_{\alpha'} - E_{\alpha}} |\alpha' \rangle \langle \alpha|
\end{eqnarray}
with the eigenstates $|\alpha\rangle,|\alpha'\rangle$ of $H_{{\rm res}} + H_{\rm Jring}$ so that 
the first-order quantities in $H_{\rm int}$ are eliminated in $H_{\rm ph}$.
The second term in Eq.~(\ref{Hph_7_21_17}) is of the second order in $H_{\rm int}$ and gives rise to the mutual couplings between resonators.

In Eq.~(\ref{Hint_7_24_17}) we approximate $n_\mu a_m$ and $n_\mu a_m^\dagger$ by
$\sum_k n_{\mu,k} |N_0,k\rangle \langle N_0,0| a_m$ and $\sum_k n_{\mu,k}^\ast |N_0,0\rangle \langle N_0,k| a_m^\dagger$, respectively, 
where $N_0$ is the relevant total charge of the Josephson ring which is conserved,
and $k(=1,2,\cdots)$ runs over the excited states of the Josephson ring, and
\begin{eqnarray}
n_{\mu,k} = \langle N_0,k| n_\mu | N_0, 0\rangle.
\label{n_11_20_17}
\end{eqnarray}
Here, $\sum_k n_{\mu,k} |N_0,k\rangle \langle N_0,0| a_m$ annihilates a photon in resonator~$m$ and yields an excitation in the Josephson ring. 
This approximation is based on the assumption that the Josephson ring remains in its ground states, and 
we neglect the counter rotating terms.
Then, the interaction Hamiltonian in Eq.~(\ref{Hint_7_20_17}) is represented as 
\begin{eqnarray}
H_{\rm int} = (2eC_c)  \sum_{m,\mu,k}  \xi_{m,\mu} V^{(m)}_{\rm rms} n_{\mu,k}
a_m  |N_0,k\rangle \langle N_0,0| + h.c..\nonumber\\
\label{Hint_7_25_17}
\end{eqnarray}
Using Eqs.~(\ref{Hph_7_21_17}) and (\ref{Hint_7_25_17}),
the effective photon lattice Hamiltonian is obtained as 
\begin{eqnarray}
H_{\rm ph} = \sum_m \varepsilon_m a_m^\dagger a_m + 
\sum_{m,n(\ne m)} t_{m,n}a_m^\dagger a_n
\end{eqnarray}
with
\begin{eqnarray}
\varepsilon_m  &=& \hbar\omega_m + (2eC_c)^2 \sum_{k>0}\Big{[} \frac{1}{\hbar\omega_m-E_k}|\Xi_{m,k} |^2 \Big{]},\nonumber\\
t_{mn}  &=& 2(eC_c)^2 \sum_{k>0}\Big{[} \Big{(} \frac{1}{\hbar\omega_m-E_k} + \frac{1}{\hbar\omega_n-E_k}\Big{)} \Xi_{m,k}^\ast  \Xi_{n,k}   \Big{]},\nonumber\\
\end{eqnarray}
where
\begin{eqnarray}
\Xi_{m,k} = V_{\rm rms}^{(m)} \sum_\mu \xi_{m,\mu} n_{\mu k}.
\end{eqnarray}

Note that $\varepsilon_m$ and $t_{m,n}$ depend on $\Phi$
through $n_{\mu,k}$ in Eq.~(\ref{n_11_20_17}).
Therefore, the coupling among resonators can be tuned {\it in-situ} via  $\Phi$.
We consider the temporal modulation of $\Phi$ around fixed value $\Phi_0$, that is, $\Phi(t)=\Phi_0+\Delta \Phi(t)$.
Then, $t_{m,n}$ and $\varepsilon_m$ are represented as
\begin{eqnarray}
t_{m,n}(\Phi) &=& t_{m,n}(\Phi_0) +  \Delta  t_{m,n}(\Delta \Phi),\nonumber\\
\varepsilon_m(\Phi) &=& \varepsilon_m(\Phi_0) +  \Delta  \varepsilon_m(\Delta \Phi).
\end{eqnarray}
We consider a narrow range of $\Delta \Phi$ in which $\Delta  t_{m,n}$ and $\Delta  \varepsilon_m$ are proportional to 
$\Delta \Phi$.
Then, the time dependence of $\Delta  t_{m,n}$ and $\Delta  \varepsilon_m$ are represented as
\begin{eqnarray}
\Delta  t_{m,n}(t) &=& \frac{d  t_{m,n}}{d \Phi}\Big{|}_{\Phi_0}   \Delta \Phi(t),\nonumber\\
\Delta  \varepsilon_m(t) &=& \frac{d  \varepsilon_m}{d \Phi}\Big{|}_{\Phi_0} \Delta \Phi(t).
\end{eqnarray}
The coupling among resonators can be tuned via the time-dependence of $\Delta \Phi(t)$.
We consider $\Delta \Phi(t)$ represented by
\begin{eqnarray}
\Delta \Phi(t) =\sum_{l} \Phi_{l} \cos(\Omega_{l}t + \theta_{l}),
\end{eqnarray}
where the index $l$ runs over every pair of the resonators, and we set $\Omega_l$ so that it matches to the difference of the resonance frequencies of resonator pair~$l$.
In the RWA the coupling strength between resonators~$m$ and~$n$ is represented as $\frac{1}{2}\frac{d  t_{m,n}}{d\Phi} \Phi_{l=(m,n)}e^{-i\theta_{l=(m,n)}}$.
We assume that the influence of the second term of $\varepsilon_m(t)$ is negligible in the RWA because it is rapidly oscillating.

Several comments are in order.
(i) Although the evaluation of $\varepsilon_m$ and $t_{m,n}$ exceeds the scope of this paper, it was reported that $|t_{m,n}|/(2\pi)$ can exceed 20~MHz for $N=3$ depending on $\Phi$~\cite{Koch2010}.
The frequency corresponding to the excitation energy from the ground state of the Josephson ring can be higher than 10~GHz.
Thus, excitation of the Josephson ring induced by oscillating $\Phi$ with frequency of orders of 100~MHz is negligible.
(ii)~$\Omega_l$ is chosen such that the oscillating $\Phi$ couples resonators.
Each pair of the resonators should have different frequency difference for selective coupling.
This limits the number of resonators.
(iii)~The electric potentials of the superconducting islands can be tuned via gate voltages to optimize 
the resonator coupling strength, although we consider the case that the gate voltages are zero for the simplicity.

\bibliography{achemso-demo}

\end{document}